\documentclass[12pt,preprint]{aastex}

\newcommand{\twelveco}{$^{12}$CO}
\newcommand{\ffactor}{0.715}
\newcommand{\emunits}{cm$^{-6}$~pc} 
\newcommand{\brgamma}{Br$\gamma$}
\newcommand{\smaunc}{15}
\newcommand{\percm}{cm$^{-1}$}
\newcommand{\persqcm}{cm$^{-2}$}
\newcommand{\percc}{cm$^{-3}$}
\newcommand{\xbandbeam}{$0\farcs60 \times 0\farcs45$} 
\newcommand{\ubandbeam}{$0\farcs79 \times 0\farcs47$} 
\newcommand{\smabeam}{$0\farcs90 \times 0\farcs79$}
\newcommand{\smapa}{-32\arcdeg}
\newcommand{\smabeamau}{$\sim$1700~AU}

\newcommand{\commonbeam}{$1\farcs40 \times 0\farcs90$} 
\newcommand{\commonpa}{+35\arcdeg} 
\newcommand{\thirtythreeso}{$^{33}$SO}
\newcommand{\thirtythreesoo}{$^{33}$SO$_2$}
\newcommand{\cthirtyfours}{C$^{34}$S}
\newcommand{\carmawave}{3~mm}

\newcommand{\kbandwave}{1.3~cm}
\newcommand{\smacontwave}{875~$\mu$m}
\newcommand{\smammwave}{1.4~mm}
\newcommand{\cseventeeno}{C$^{17}$O}
\newcommand{\soo}{SO$_2$} 
\newcommand{\hthirteencoplus}{H$^{13}$CO$^+$}
\newcommand{\thirtyfoursoo}{$^{34}$SO$_2$}
\newcommand{\hcccn}{HC$_3$N}
\newcommand{\hcccnv}{HC$_3$N~$v_7=1$}
\newcommand{\soov}{SO$_2$~$v_2=1$}
\newcommand{\kms}{km~s$^{-1}$}
\newcommand{\jkms}{Jy~beam$^{-1}$~km~s$^{-1}$}
\newcommand{\jyb}{Jy~beam$^{-1}$}
\newcommand{\gfive}{G5.89-0.39}

\newcommand{\water}{H$_2$O} 
\newcommand{\htwos}{H$_2$S} 
\newcommand{\hcoplus}{HCO$^+$} 
\newcommand{\chthreecn}{CH$_3$CN} 
\newcommand{\methanol}{CH$_3$OH}
\newcommand{\htwodplus}{H$_2$D$^+$}

\newcommand{\ammonia}{NH$_3$}
\newcommand{\ntwohplus}{N$_2$H$^+$}
\newcommand{\lsun}{L$_\odot$}
\newcommand{\msun}{M$_\odot$}
\newcommand{\mjb}{mJy~beam$^{-1}$}
\newcommand{\jb}{Jy~beam$^{-1}$}

\newcommand{\mirlambda}{$11.9~\mu$m}
\newcommand{\nirlambda}{$2.12~\mu$m}

\begin{document}

\shortauthors{Hunter et al.}

% 44 character maximum
\shorttitle{Submillimeter Imaging of G5.89-0.39}

\title{Subarcsecond Submillimeter Imaging of the Ultracompact HII Region G5.89-0.39}

\author{
  T.R. Hunter\altaffilmark{1},
  C.L. Brogan\altaffilmark{1}, 
  R. Indebetouw\altaffilmark{1,2},
  C.J. Cyganowski\altaffilmark{3}
}
 
\email{thunter@nrao.edu}
\email{cbrogan@nrao.edu}
\email{remy@virginia.edu}
\email{ccyganow@astro.wisc.edu}

\altaffiltext{1}{NRAO, 520 Edgemont Rd, Charlottesville, VA, 22903} 
\altaffiltext{2}{University of Virginia, Astronomy Dept., P.O. Box 3818, Charlottesville, VA, 22903-0818 } 
\altaffiltext{3}{University of Wisconsin, Madison, WI 53706} 

%\slugcomment{submitted to {\it The Astrophysical Journal}}

\begin{abstract}

We present the first subarcsecond submillimeter images of the
enigmatic ultracompact HII region (UCHII) G5.89-0.39.  Observed with
the SMA, the 875~$\mu$m continuum emission exhibits a shell-like
morphology similar to longer wavelengths.  By using images with
comparable angular resolution at five frequencies obtained from the
VLA archive and CARMA, we have removed the free-free component from
the 875~$\mu$m image. We find five sources of dust emission: two
compact warm objects (SMA1 and SMA2) along the periphery of the shell,
and three additional regions further out.  There is no dust emission
inside the shell, supporting the picture of a dust-free cavity
surrounded by high density gas.  At subarcsecond resolution, most of
the molecular gas tracers encircle the UCHII region and appear to
constrain its expansion.  We also find G5.89-0.39 to be almost
completely lacking in organic molecular line emission. The dust cores
SMA1 and SMA2 exhibit compact spatial peaks in optically-thin gas
tracers (e.g. $^{34}$SO$_2$), while SMA1 also coincides with
11.9$\mu$m emission.  In CO(3-2), we find a high-velocity north/south
bipolar outflow centered on SMA1, aligned with infrared H$_2$ knots,
and responsible for much of the maser activity.  We conclude that SMA1
is an embedded intermediate mass protostar with an estimated
luminosity of 3000~\lsun\ and a circumstellar mass of $\approx
1$~\msun.  Finally, we have discovered an NH$_3$ (3,3) maser
12\arcsec\ northwest of the UCHII region, coincident with a 44~GHz
CH$_3$OH maser, and possibly associated with the Br$\gamma$ outflow
source identified by \citet{Puga06}.

\end{abstract}


\section{Introduction}

The formation of massive stars is a fundamental process in
astrophysics that is not well understood.  The identification of
ultracompact HII regions (UCHIIs) by \citet{Wood89} provided an
important step forward in the study of massive star formation.  These
objects are less than $10^{17}$~cm in diameter, have a centimeter
wavelength spectral index consistent with free-free emission, and are
associated with far-infrared and submillimeter sources of high
bolometric luminosity \citep{Hunter00,Thompson06}.  Based on these
characteristics, UCHIIs are believed to be ionized nebulae powered by
young OB stars located at (or near) their centers and surrounded by
high column densities of dust and molecular gas \citep{Churchwell02}.
However, because of the high extinction at optical and infrared
wavelengths, the exciting star is often too deeply embedded for its
photospheric emission to be observed.  In fact, until recently the
exciting stars had been identified for only a few percent of UCHII
regions \citep{Hanson02}, with the first (and perhaps best) example
being G29.96-0.02 \citep{Watson97,MartinHernandez03}.  Recent infrared
studies have identified candidate ionizing sources toward several more
UCHIIs \citep{Alvarez04,Apai05,Comeron06,Bik06}.

One UCHII for which a candidate ionizing star has been proposed is
W28~A2, also known as \gfive, which is a well-known shell-type UCHII.
\citet{Feldt03} have identified a near-infrared star located along the
northeastern rim of the radio shell which they argue is of spectral
type O5 or earlier (hereafter called Feldt's star).  \gfive\ is
associated with a strong bipolar molecular outflow originally
identified in CO by \citet{Harvey88} and having one of the largest
mass outflow rates known \citep{Churchwell97}.  The outflow emission
has been subsequently studied using single-dish CO and SiO
\citep{Klaassen06,Choi93,Acord97} and interferometric CO, \hcoplus\
\citep{Watson07} and SiO \citep{Acord97,Sollins04} observations.  The
reported position angle of the outflow is notably different between
the tracers, being nearly east-west in CO and \hcoplus\ (position
angle +84\arcdeg), vs. northeast-southwest in SiO (position angle
+30\arcdeg).  While masers of many kinds are seen toward and around
\gfive\ (the most recent images being: \citet{Hofner96} (\water),
\citet{Stark07} and \citet{Fish05} (OH), and \citet{Kurtz04} (Class I
\methanol)), the spatial-velocity gradients seen do not lead to a
clear consensus view on the origin of the outflow.  Furthermore,
despite the fact that the purported O5 star lies along the axis of the
SiO velocity gradient, it does not reside midway between the red and
blue peaks, nor does it reside at the center of the radio shell.  Such
an asymmetry of the powering star's location with respect to the radio
shell was predicted by \citet{Ball92} in their attempt to reconcile
the mid-infrared and radio morphologies.  Their model predicts the
presence of dense molecular gas close to the northeast edge of the
UCHII.  While single-dish mm/submm surveys have established that
\gfive\ is a molecular line-rich source in the \smacontwave\ window
\citep{Hatchell98}, the majority of the lines are from sulfur-bearing
species rather than the heavy organic molecules more typically seen in
hot cores \citep{Thompson99}.  Clearly, higher angular resolution
studies of molecular gas and dust are required to test the various
hypotheses on the nature of this complex massive star formation region
and attempt to unify the observed phenomena.

Toward this end, we have obtained the first subarcsecond submillimeter
(\smacontwave) observations of \gfive.  Our Submillimeter Array (SMA)
images have a factor of seven higher angular resolution (in beam area)
than the \smammwave\ \citet{Sollins04} data, and are substantially
more sensitive to dust emission and molecular lines.  We interpret our
unprecedented submillimeter images in the context of new images with
comparable resolution constructed from the best available centimeter
wavelength data in the VLA archive, 3~mm data from the new
B-configuration of the Combined Array for Millimeter Astronomy
(CARMA), and archival near- to mid-infrared data.

The distance to \gfive\ used in the literature varies, although all
papers agree that the source does not lie at the far kinematic
distance.  \citet{Acord98} report a detection of the expansion of the
nebula (though with a fairly low signal-to-noise of $2.5\pm0.5$
mas~yr$^{-1}$) in multiepoch VLA images, which implies a distance of
2.0$^{+0.7}_{-0.4}$~kpc.  The kinematic distance based on the
main-line OH maser velocity of +18~\kms\ is $3.8^{+1.0}_{-1.7}$~kpc
\citep{Fish03}.  The kinematic distance based on the +9.3~\kms\ LSR
velocity of CS(2-1) \citep{Bronfman96} is 2.6~kpc, while the distance
to the nearby W28 supernova remnant is 1.9~kpc \citep{Velazquez02}.
We will adopt a distance of 2.0~kpc in this paper.

\section{Observations}

\subsection{Submillimeter Array (SMA)}

% dopplerTrack -r 347.33063 -u -s23  = LO ~ 341.46

The SMA\footnote{The Submillimeter Array (SMA) is a collaborative
project between the Smithsonian Astrophysical Observatory and the
Academia Sinica Institute of Astronomy \& Astrophysics of Taiwan and
is funded by the Smithsonian Institution and the Academia Sinica.}
observations were made with eight antennas in the extended
configuration on 2006 June 05.  Sky conditions were clear and the
225~GHz opacity measured by the tipping radiometer at the Caltech
Submillimeter Observatory (CSO) varied from 0.065 to 0.081 during the
observations.  The SMA receivers have double sideband mixers with 2~GHz
bandwidth centered at an intermediate frequency of 5~GHz.  The
front-end local oscillator was tuned to place the SiO(8-7) spectral
line ($\nu=347.33063$ GHz) in the center of correlator chunk 23 in the
upper sideband (USB), resulting in center frequencies of 336.46~GHz in
lower sideband (LSB) and 346.46~GHz in USB.  The data were taken with
a channel width of 0.81~MHz (0.71 \kms), but were subsequently Hanning
smoothed for a final spectral resolution of 1.4 \kms.  The phase
center was $18^{\rm h}00^{\rm m}30^{\rm s}.45$, $-24^{\circ}04'01''.5$
(J2000), and the projected baseline lengths ranged from 20 to 230
k$\lambda$.  Typical system temperatures were 200 to 300~K at
transit and the total time on-source was 4 hours. The gain calibrators
were NRAO530 (J1733-130) (13\arcdeg\ distant) and J1911-210
(17\arcdeg\ distant) and the bandpass calibrator was J2232+117.  Flux
calibration is based on observations of Ceres and a model of its
brightness distribution using the Miriad task smaflux. Comparison of
the derived flux of the quasars with SMA flux monitoring suggests the
absolute flux calibration is good to within \smaunc\%.  The estimated
accuracy of the absolute coordinates is $\sim 0.15$\arcsec. The uncertainty
in the relative positions is better than $\sim 0.01$\arcsec.

The data were calibrated in Miriad, then exported to AIPS where the
line and continuum emission were separated with the task UVLSF (using
only line-free channels to estimate the continuum).  Self-calibration
was performed on the continuum data, and solutions were transferred to
the line data.  The continuum and line data were imaged using natural
weighting. All channels were cleaned to a flux density limit of
350~\mjb\/, and individual data cubes were then extracted for each
spectral line detected.  After combining the calibrated LSB and USB
continuum uv-data, the 1$\sigma$ rms noise level achieved in the
continuum image is 6.4~\mjb.  The noise level in a single channel of
the spectral line images is 110~\mjb.  The line data have been
corrected for the half channel error in SMA velocity labeling
discovered in November 2007. The synthesized beam is \smabeam\ at
position angle \smapa, corresponding to a linear scale of \smabeamau.
The primary beam is $\sim 37\arcsec$; these data are insensitive to
smooth structures larger than about 10\arcsec.

We also recalibrated and imaged the \smammwave\ data from the SMA
compact configuration originally published by \citet{Sollins04} (the
details of the observing setup can be found in that paper). Using
natural weighting, we achieved a continuum rms of 24~\mjb\ and a
spectral line rms of 150~\mjb\ (per 1.1~\kms\ channel), with a
beamsize of $3\farcs1 \times 2\farcs6$ (P.A.=5.5\arcdeg).  The imaged
spectral lines include SiO (5-4), \htwos\ ($2_{2,0}-2_{1,1}$), and
\hcccn\ (25-24).

\subsection{James Clerk Maxwell Telescope (JCMT)}

We have obtained newly-processed 850~$\mu$m SCUBA imaging data of
\gfive\ observed at the JCMT\footnote{ The James Clerk Maxwell
Telescope is operated by the Joint Astronomy Centre on behalf of the
Science and Technology Facilities Council of the United Kingdom, the
Netherlands Organisation for Scientific Research, and the National
Research Council of Canada.} on Mauna Kea, Hawaii.  The data were
obtained from the Canadian Astronomy Data Center (CADC) repository of
SCUBA Legacy Fundamental Object
Catalogue.\footnote{http://www4.cadc-ccda.hia-iha.nrc-cnrc.gc.ca/community/scubalegacy/}
The image of the source was constructed from a number of independent
observations and has an effective beamsize of $22\farcs9$.  Further
information on the data processing is described by
\citet{DiFrancesco08}.  The total flux density of the source was
determined from the image to be 36~Jy.  For comparison, the flux
density reported in the table of the SCUBA Legacy Catalogue is also
36~Jy \citep[see also][]{Sandell94}.  Due to the nature of the chopped
observations, these values should be considered lower limits.  In
order to be conservative, we have added 36~Jy as the zero spacing flux
density to the SMA uv data.  With this addition, the total flux
density recovered in the high resolution image is 9.6~Jy, or 27\%.

\subsection{Very Large Array (VLA)}

Several archival datasets from the NRAO\footnote{The National Radio
Astronomy Observatory is a facility of the National Science Foundation
operated under agreement by the Associated Universities, Inc.} Very
Large Array (VLA) were calibrated and imaged in AIPS.  The observing
parameters of these data are summarized in Table~\ref{vlaobs}.  Some
of the datasets were observed in continuum mode, while others were
observed in line mode.  In project AZ075, the \ammonia\ (2,2) and
(3,3) transitions were observed simultaneously, each with 6.3~MHz
bandwidth (82 \kms) with 0.195~MHz channel width (2.45 \kms).

\subsection{Combined Array for Millimeter Astronomy (CARMA)}

Our CARMA\footnote{Support for CARMA construction was derived from the
states of California, Illinois, and Maryland, the Gordon and Betty
Moore Foundation, the Kenneth T. and Eileen L. Norris Foundation, the
Associates of the California Institute of Technology, and the National
Science Foundation.  Ongoing CARMA development and operations are
supported by the National Science Foundation under a cooperative
agreement, and by the CARMA partner universities.} observations were
obtained on 12 December 2007 in the B-configuration with 13 antennas
(five 10.4m antennas and eight 6m antennas, for a total of 78
baselines).  The 225~GHz zenith opacity ranged from $\sim$0.19 to 0.23
during the observations.  Typical (SSB) system temperatures were 200
to 400~K and the total time on-source was 1.8 hours.  The receivers
were tuned to an LO frequency of 92.85~GHz.  Three continuum bands of
$\sim$500~MHz width, each comprised of 15 channels, were correlated in
each sideband, centered at $\sim$89.6, 90.7, and 91.2 ~GHz (LSB) and
94.5, 95.0, and 96.1 ~GHz (USB).  Only the bands centered at 90.7,
91.2, 94.5, and 95.0 ~GHz were used due to the significantly reduced
sensitivity on the 6m antennas at IFs above 2.4~GHz.  The phase center
was $18^{\rm h}00^{\rm m}30^{\rm s}.45$, $-24^{\circ}04'01''.5$
(J2000). The gain calibrator was NRAO530 (J1733-130) (13\arcdeg\
distant), the bandpass calibrator was 3C454.3, and the flux calibrator
was Uranus.  The data were calibrated in Miriad and then
self-calibrated in AIPS.  The combination of the LSB and USB data
imaged with a robust weighting factor of -0.5 yielded a beamsize of
$1\farcs40\times0\farcs90$ at position angle 38\arcdeg, and an image
rms of 1.6~\mjb.

\subsection{Infrared images}

Several archival images were obtained from the ESO archive, as listed
in Table~\ref{esodata}.  In particular, the data include those
published in \citet{Feldt03} from which those authors identified the
proposed exciting source which we call "Feldt's star".  The
near-infrared ($\lambda<$5~$\mu$m) images were processed with standard
techniques: each raw frame was reprojected to a common projection, and
the raw frames combined with a median filter to remove outliers and
cosmic rays.  No sky subtraction was attempted, since we use these
images to identify infrared sources and morphology, but not for
photometric measurements. Although the relative pointing of raw frames
was very accurate, the absolute astrometry of the data were incorrect
by up to 10\arcsec. We registered the relatively large field-of-view
ISAAC image to 2MASS; since the latter has astrometric uncertainty of
less than 0.1\arcsec, this registration is limited by centroiding
the $\sim$10 stars used, with a resulting uncertainty on the ISAAC
frame position of 0.1\arcsec.  We then registered the smaller
field-of-view NACO frames to the ISAAC image using 5--10 stars in each
frame, with a total resulting positional uncertainty of 0.2\arcsec.
Our positions differ from those of \citet{Feldt03} by 0.2\arcsec\ (our
position is nearly directly east of theirs).  However, since the
offset we find is equal to the uncertainty, for this paper we have
chosen to align our near-IR images so that the position of Feldt's
star matches the value published by those authors (18:00:30.44,
-24:04:00.9 (J2000)).

The mid-infrared TIMMI2 data required somewhat more detailed
processing to remove slowly-changing flat field and detector effects
over the course of the observation\footnote{See the TIMMI2
documentation
\url{http://www.ls.eso.org/lasilla/sciops/3p6/timmi/html/t2\_overview.html}
for more information.}.  We constructed a "rolling sky" sequence of
frames by taking the median of each four frames in chronological
order.  As the raw frames are chopped and the telescope nodded, each
frame consists of a positive sky image minus a shifted negative sky
image, with large-scale thermal emission removed.  The position of the
source changes according to the telescope nod position, so the median
of four consecutive frames contains no information from the sky, but
information about time-dependent detector variations. The rolling sky
is subtracted from each raw frame, those are masked, shifted, and
added to form a mosaic image.  Finally, that image had to be
registered, which was done using the astrometry of the mid-infrared
images in \citet{Feldt99} since we could not find the Brooks image
(Table~\ref{esodata}) in any publication.

\section{Results} 

\subsection{Continuum images}
\label{continuum}

\begin{figure}
\epsscale{1.0}
% generated by      plots/smacontwith220u.greg
\includegraphics[width=6.5in,angle=0]{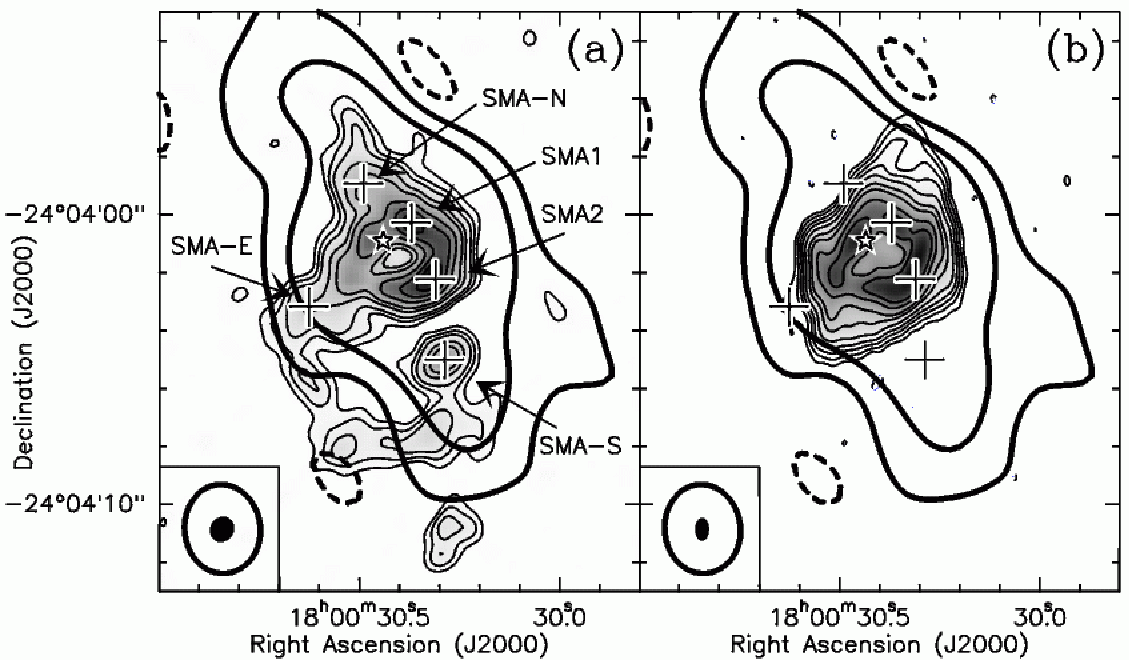} 
\caption{(a) Greyscale and thin contours: SMA continuum image
at \smacontwave\ with a resolution of \smabeam\ at position angle
\smapa. Contour levels are -20, 20, 39, 65, 130, 195, 325, 455,
585 \mjb.  Thick contours: SMA continuum image at \smammwave\ with a
resolution of $3\farcs1 \times 2\farcs6$ at position angle +5\arcdeg.
Contour levels are -50, 50, 250 \mjb.  (b) Greyscale and thin
contours: VLA 2~cm image with a resolution of \ubandbeam\ at
position angle -1\arcdeg. 
%xband Contour levels are -0.5, 0.5, 1.7, 4.2,
%8.3, 16.7, 33.3, 50, 83.3, 100, 117 \mjb.  
Contour levels are -0.9, 0.9, 3, 7.5, 15, 30, 60, 90, 150, 210 \mjb.
Thick contours: SMA continuum image at \smammwave.  The respective
beams are shown in the lower left corners. The star symbol marks the
location of Feldt's star \citep{Feldt03}.  The large crosses denote
regions of excess 875~$\mu$m emission above that expected from
free-free emission extrapolated from cm wavelengths (see
Figure~\ref{cont4panel} and Table~\ref{dustcores}).
\label{smacontsollins}
}
\end{figure}

The subarcsecond resolution SMA~\smacontwave\ continuum image is shown
in Figure~\ref{smacontsollins}a.  For comparison, we overlay two
contours from the lower resolution \smammwave\ image.  The north-south
extension first noted by \citet{Sollins04} is now resolved into
discrete sources at \smacontwave; these regions have been labeled for
later reference.  The submillimeter emission shows a shell-like
morphology with an obvious hole in the center. The brightness of the
shell varies around its circumference with three major peaks.  The
easternmost peak lies $0\farcs25$ east of Feldt's star.  Some emission
(SMA-N) extends off the north-northeast side of the shell, and a point
source (SMA-S) is seen completely separate from the shell to the
south-southwest.  The point of lowest brightness in the UCHII shell is
on the southeast side and may indicate an opening in the shell.
Further to the south-southeast, we find faint filamentary emission
that appears to trace a coherent structure.  No \smacontwave\
continuum emission was found outside the field shown in
Figure~\ref{smacontsollins}a.

Our 2~cm image is shown in Fig~\ref{smacontsollins}b. This is the
highest resolution and sensitivity image to date of \gfive\ at radio
wavelengths shorter than 3.6~cm.  As seen in the original 2~cm survey
image of \citet{Wood89}, the dominant feature is a shell-like
structure surrounding a central cavity of diameter $\approx 1$\arcsec.
The elongated low-level emission surrounding the shell evident along a
position angle of $\approx -28$\arcdeg\ has been known since the
original 6~cm VLA A-configuration image of \citet{Zijlstra88}. In our
new image, we find a nearly-closed loop of emission extending along
this direction on the northwest side of the shell \citep[also
see][]{Acord98}.  Like the \smacontwave\/ image, the 2~cm image shows
a break in the SE edge of the shell, suggesting a ''blow-out'' of
ionized gas in this direction.  We note that the southern portion of
the elongated ionized emission is not exactly coincident with the
extension seen at \smacontwave, but instead lies to its SW.

\begin{figure}
% generated by 3MMCON/allcon9carma.greg
\includegraphics[width=6.5in,angle=0]{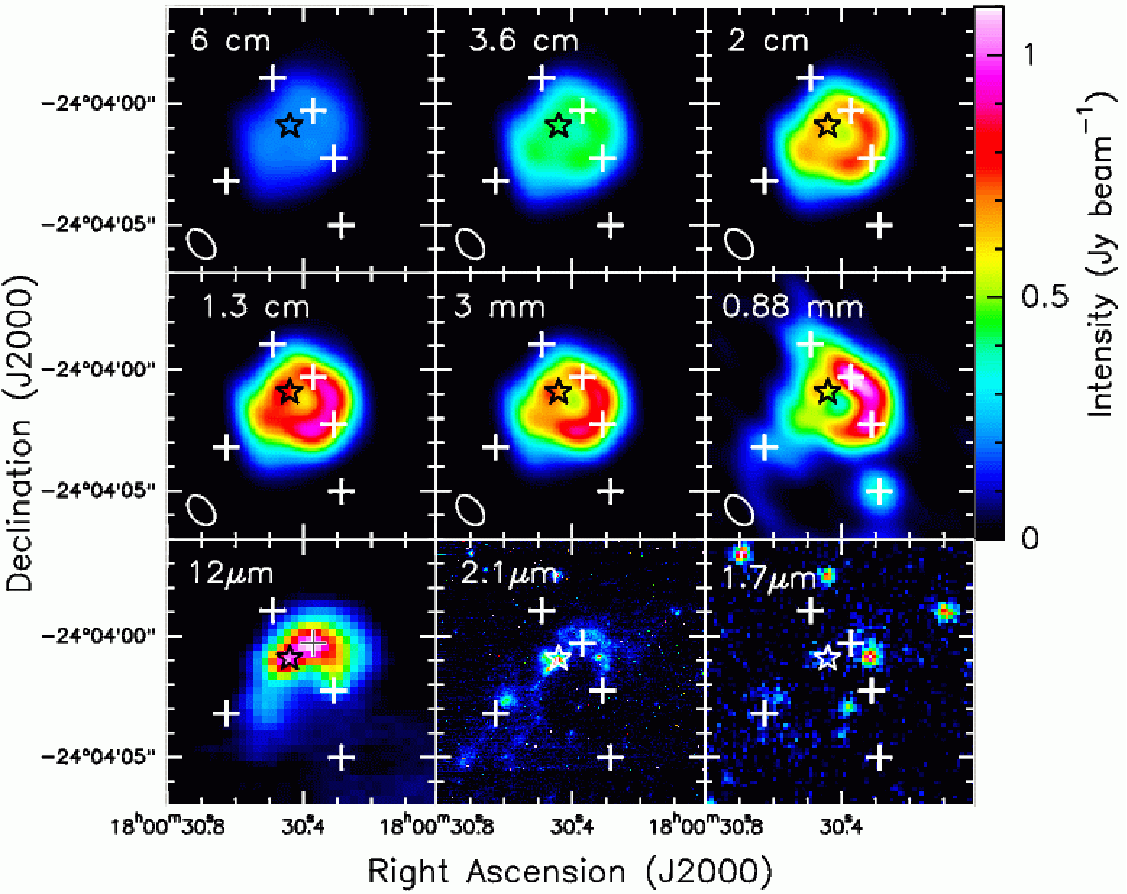} % {f2.ps}
\caption{Continuum images of \gfive\ from low frequency in the upper
left to high frequency in the lower right.  The top six images have
been restored with the beamsize of the 3~mm image (\commonbeam\ 
at P.A.=\commonpa), and are
shown on the same colorscale (0.0 to 1.1 \jb). The white crosses mark
the positions of the submillimeter dust sources identified in
Figure~\ref{cont4panel} and Table~\ref{dustcores}.   
The star symbol marks the position of Feldt's star \citep{Feldt03}.   
\label{vlacont}}
\end{figure}

For comparison, the SMA \smacontwave\ and VLA 2~cm images are shown
alongside a wide range of other centimeter to near-infrared wavelength
images in Figure~\ref{vlacont}.  All of the radio through
submillimeter images have been generated from data with a comparable
range of uv spacings, have been restored with the same beam
(\commonbeam\ at P.A.=\commonpa), and are displayed with the same flux
density color scale.  The beam was chosen to match the beam of the
poorest resolution dataset (3~mm).  At the lowest frequency (6~cm),
the central part of the UCHII region shows a roughly uniform
appearance.  As one moves to higher frequency, the source brightens,
and the shell structure emerges and becomes increasingly distinct.
This change in appearance is due to the decreasing optical depth of
free-free emission as the observed frequency increases.

\begin{figure}
% generated by plots/carma.greg
\includegraphics[width=5.5in,angle=0]{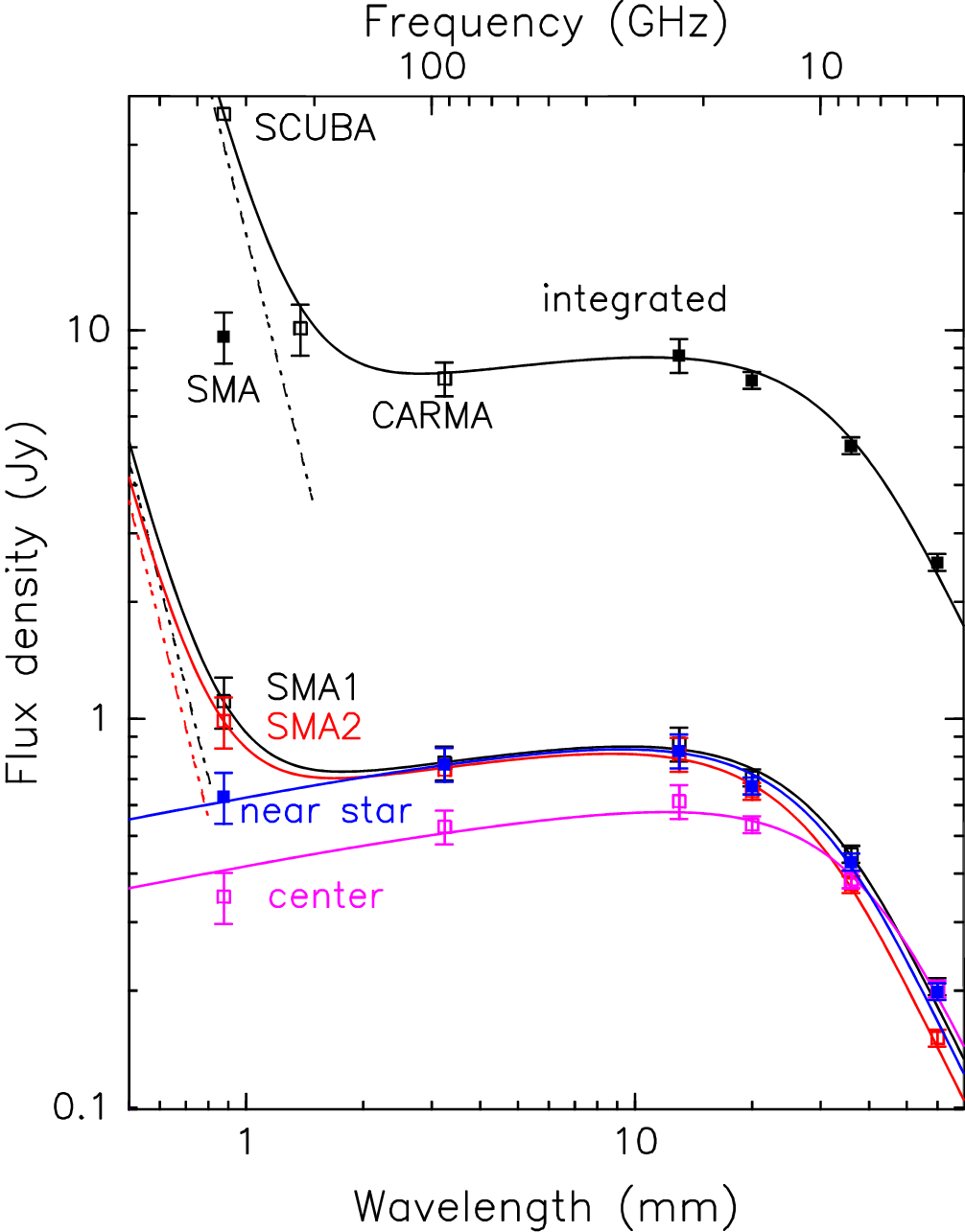}
\caption{ Spectral energy distribution at four positions.  Lower black
line and points: SMA1; red line and points: SMA2; purple line and
points: center of shell; blue line and points: $0\farcs3$ east of
Feldt's star (see Table~\ref{cmflux}).  The upper black line and points
are the flux densities integrated over a 12\arcsec\ square box.  The
solid lines are models constructed from free-free emission
\citep{Beckert00}, plus (in the case of SMA1, SMA2 and integrated)
dust emission with a $\nu^4$ spectrum, shown as a dash-dot line.  The
upper open squares are the SCUBA peak flux density and the integrated
flux densities from the \smammwave\ SMA data and the \carmawave\ CARMA data.
The error bars reflect the uncertainty values in Table~\ref{cmflux}.
\label{freefreesed}}
\end{figure}

To further demonstrate the effect of free-free optical depth, we
measured the flux density in a single beam at various positions in the
UCHII region (using the images in Figure~\ref{vlacont}) as well as the
total emission in the field within an 12\arcsec\ box (see
Table~\ref{cmflux}).  The corresponding spectral energy distributions
(SEDs) are plotted in Figure~\ref{freefreesed}.  At centimeter to
millimeter wavelengths, the data are consistent with free-free
emission with a turnover at about 2~cm.  The excess emission at
\smacontwave\ toward SMA1 and SMA2 suggests the presence of dust
toward these positions. In order to quantify this component, we must
estimate and remove the free-free contribution to the \smacontwave\
emission.  The common analytic approximations for free-free emission
given by \citet{Altenhoff60} and \citet{Scheuer60} become less
accurate above 100~GHz.  Thus, we have used the numerical calculation
of the Gaunt factor for free-free radiation along with Equations~2 and
15 given by \citet{Beckert00}.  Using the non-linear fitting utility
of Matlab (lsqcurvefit), we produced the SED fits shown in
Figure~\ref{freefreesed}.  The resulting electron temperatures and
emission measures range from 9000~K and $3.0 \times 10^8$~\emunits\
at the central position to 12000~K and $8.3 \times 10^8$~\emunits\
close to Feldt's star.  For the integrated emission, the corresponding
values are 8500~K and $3.5 \times 10^8$~\emunits.  Assuming the source
has equal extent in the third dimension as it does in the plane of the
sky (4\arcsec\ or 0.039~pc) this average emission measure corresponds
to an average electron density of $\sim 10^5$~\percc. However, this
estimate does not take into account the shell-like structure of the
UCHII, and in fact the width of the shell is only marginally resolved
with $0\farcs7$ resolution, implying that significantly higher $n_e$
are present in the shell.

\begin{figure}
\epsscale{1.0}
% generated by   plots/cont4panel.greg
\includegraphics[height=5.5in,angle=0]{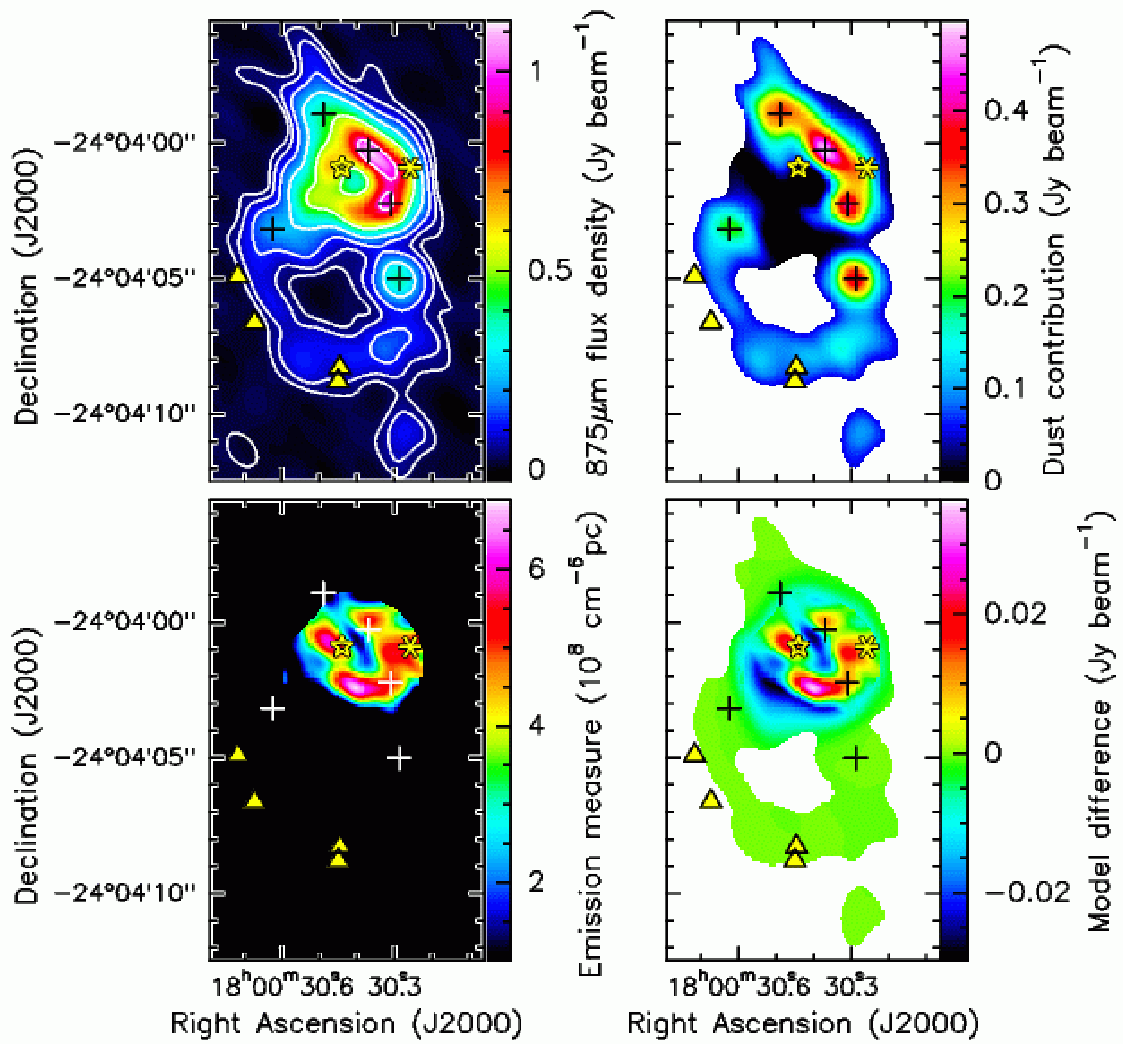} % {f4.ps} 
\caption{{\it Top left panel:} Observed SMA continuum image at
\smacontwave\ restored with a resolution of \commonbeam\ at position
angle \commonpa. The beam is shown in the lower left corner.  Contour
levels are -0.03, 0.06, 0.12, 0.24, 0.48, 0.96 \jb.  {\it Top right
panel:} Estimated dust contribution to the observed flux density,
computed by scaling the \kbandwave\ image by \ffactor, and removing it
from the \smacontwave\ image.  The \smacontwave\ image was masked at
the 50~mJy level prior to subtraction.  {\it Bottom left panel:} Model
of the emission measure vs. position computed by comparing the flux
ratio between the \kbandwave\ and \carmawave\ images on a
pixel-by-pixel basis. At 1.3~cm, the model has a mean optical depth of
0.115, and a maximum optical depth of 0.268.  {\it Bottom right
panel:} The difference between the dust image shown in the top right
panel and the dust image produced by using the free-free model shown
in the bottom left panel.  In all panels, the star symbol marks the
location of Feldt's star, and the black crosses denote the positions
of sources of dust emission (see Table~\ref{dustcores}). The triangles
mark near-IR H$_2$ emission components (A, D1, D2), and the yellow
asterisk is the \brgamma\ outflow origin, all from \citet{Puga06}.
\label{cont4panel} 
}
\end{figure}

Based on these results, we followed two methods for deriving a
free-free emission map at \smacontwave\ and compared their results.
The first and simplest method consisted of choosing a characteristic
temperature (10000~K) and emission measure ($4 \times 10^8$~\emunits),
computing the expected flux density scale factor between \kbandwave\
and \smacontwave, and applying it uniformly to all pixels in the
\kbandwave\ image.  The scale factor (\ffactor) corresponds to an
effective spectral index of -0.154 from \kbandwave\ to
\smacontwave\footnote{While it should be noted that the scale factor
determined in this manner is slightly dependent on the electron
temperature, it changes by $<0.1$\% for temperatures in the range of
7500-15000~K.}.  This spectral index is in good agreement with the
values used in recent attempts to quantify the Galactic foreground
contribution to CMB studies \citep[see for example][]{Dickinson03}.
The extrapolated free-free image was then subtracted from the observed
\smacontwave\ image in order to generate an image of dust emission,
shown in Figure~\ref{cont4panel}.  Due to the subtraction, the image
rms is increased by $\approx \sqrt{2}$ to 10~\mjb, thus it has been
masked at the $5\sigma$ level (50~\mjb).

For the second method, we first computed the ratio of the \kbandwave\
and \carmawave\ images, then solved for the emission measure as a
function of pixel that would theoretically produce the observed
ratio, assuming isothermal 10000~K gas.  The result is an emission
measure image (shown in Figure~\ref{cont4panel}), which we then used
to generate a synthetic free-free image at \smacontwave, which we
finally removed from the observed \smacontwave\ image.  The resulting
dust image closely matched that generated by the simpler method, as
seen in the ``model difference'' image in Figure~\ref{cont4panel}.
The largest differences are less than $5\sigma$ when compared with the
rms in the observed \smacontwave\ image.  

Five discrete areas of dust emission are evident, and their positions
and flux densities are tabulated in Table~\ref{dustcores}.  SMA1 and
SMA2 are compact sources which lie on or near the UCHII shell, SMA-S
is compact and lies outside of the shell, and SMA-N and SMA-E are
more extended and also lie outside of the shell.  As can be seen in
the VLA images (Figure~\ref{vlacont}), there is no corresponding
centimeter wavelength emission toward the positions denoted SMA-N,
SMA-S, and SMA-E; thus we conclude that this emission is from dust.
We note that the locations of SMA-N and SMA-S lie along an axis
defined by the southern cluster of OH masers \citep{Stark07} and
passing through the center of the UCHII shell.  Further south, the
filamentary structure of \smacontwave\ emission must also arise from
dust as there is no evidence for it in the centimeter images, which
have more than adequate sensitivity to have detected it (if it was
free-free emission).  We note that there is no evidence for dust
emission near the center of the UCHII shell.  The \smacontwave\ flux
density from dust at the position of Feldt's star is 10~\mjb\ which is
equal to the $1\sigma$ noise level.  For comparison, if we use our
astrometry for Feldt's star, the flux density from dust would be
31~\mjb.

\subsection{Infrared images}

Images at 1.7~$\mu$m, \nirlambda\ ($K_s$ band) and \mirlambda\ ($N$
band) are shown in the bottom row of Figure~\ref{vlacont}; Feldt's
star is indicated by the open star symbol.  The cometary shape of the
mid-infrared emission toward \gfive, coinciding with the northern half
of the UCHII shell, has been noted previously \citep{Ball92}. However,
the $N$ band image also shows evidence for two distinct peaks within
the cometary emission.  The position of the brighter $N$ band peak is
coincident with that of SMA1 (i.e. within the combined astrometric
uncertainties).  Similarly, the fainter peak coincides with Feldt's
star.  In agreement with \citet{Puga06}, we find that most of the
$K_s$ band emission in the field shown in Figure~\ref{vlacont} is
continuum rather than H$_2$, with the exception of the sources A, D1,
and D2 denoted by \citet{Puga06}.  We also examined {\em Spitzer} IRAC
data for this region but with its poorer angular resolution
($>2$\arcsec) the mid-IR emission appears as a featureless
blob. Comparison of the narrow band 1.2~$\mu$m image (not shown, see
Table~\ref{esodata}) with the 1.7~$\mu$m image shows that all four of
the bright stars in the upper half of the 1.7~$\mu$m image shown in
Figure~\ref{vlacont} are also detected at 1.2~$\mu$m, with the star
nearest SMA1 being the faintest.  It is likely that all four stars are
foreground to the nebula.
%
%The N band flux densities are listed in Table~\ref{nband}.

\subsection{Spectral line images} 
\label{s1}

\begin{figure}
% generated by   plots/mol9port.greg
\includegraphics[width=6.5in]{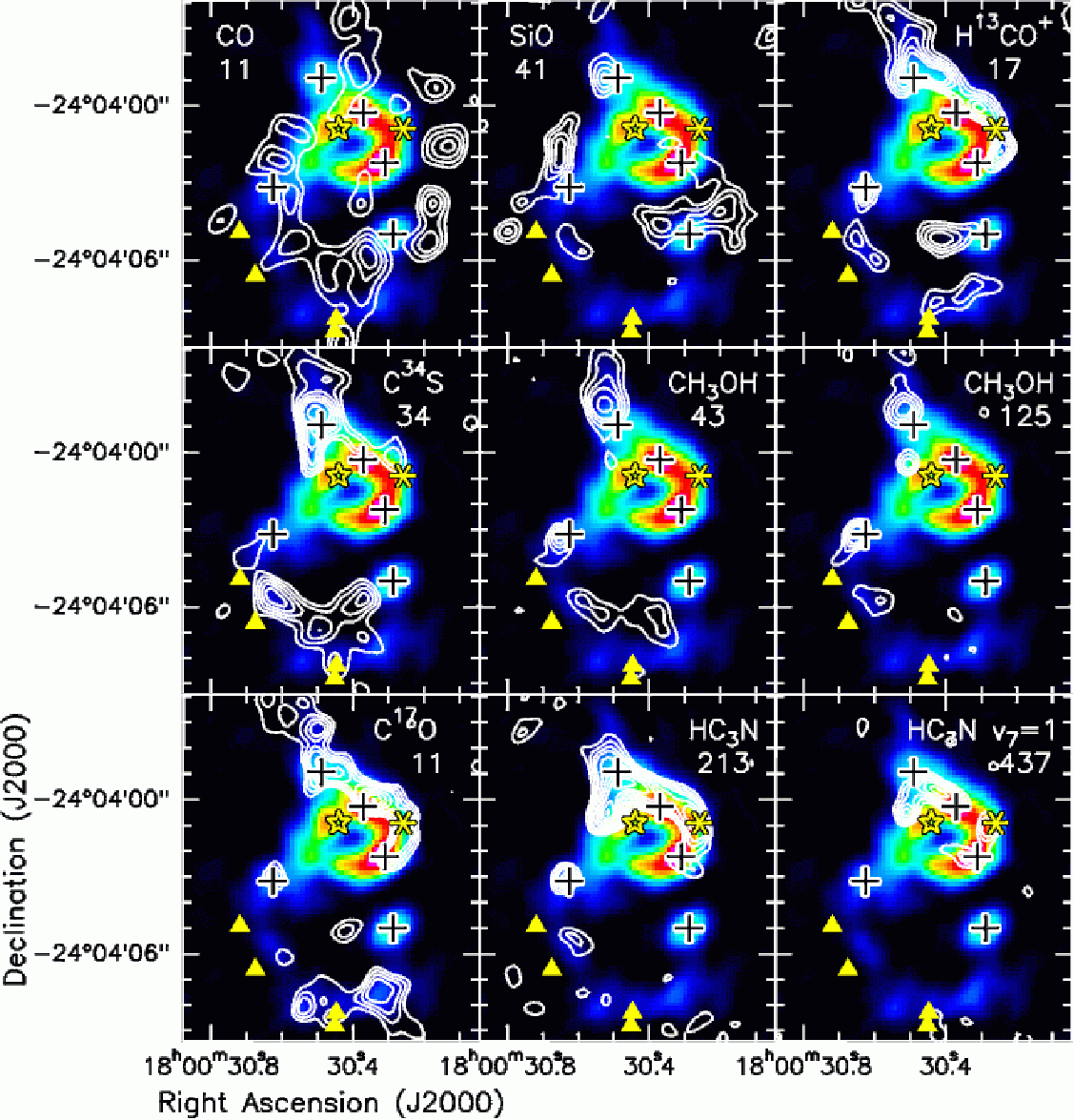}
\caption{Integrated intensity contour maps of submillimeter spectral 
lines overlaid on the
SMA \smacontwave\ continuum image.  The black crosses mark the positions 
of the dust sources SMA1, SMA2, SMA-N, SMA-E, and SMA-S (see
Table~\ref{dustcores}).  The yellow triangles mark the near-IR H$_2$
knots (A, D1, D2), and the yellow asterisk marks the origin of
the \brgamma\ outflow \citep{Puga06}.  Contour levels are 3, 4, 5, 6, 7,
8, 10, 12, 14, 16, 18, 20 $\times$ the base level listed in
Table~\ref{smalines}. The number below the species name is the energy
above ground for the lower level in units of \percm.  The image for
\hcccn\ is the average image of the two lines in the table.  The image
for \hcccnv\ is the average image of the three lines in the table.
\label{mol9}}
\end{figure}

\begin{figure}
% generated by   plots/mol9portso.greg
\includegraphics[width=6.5in]{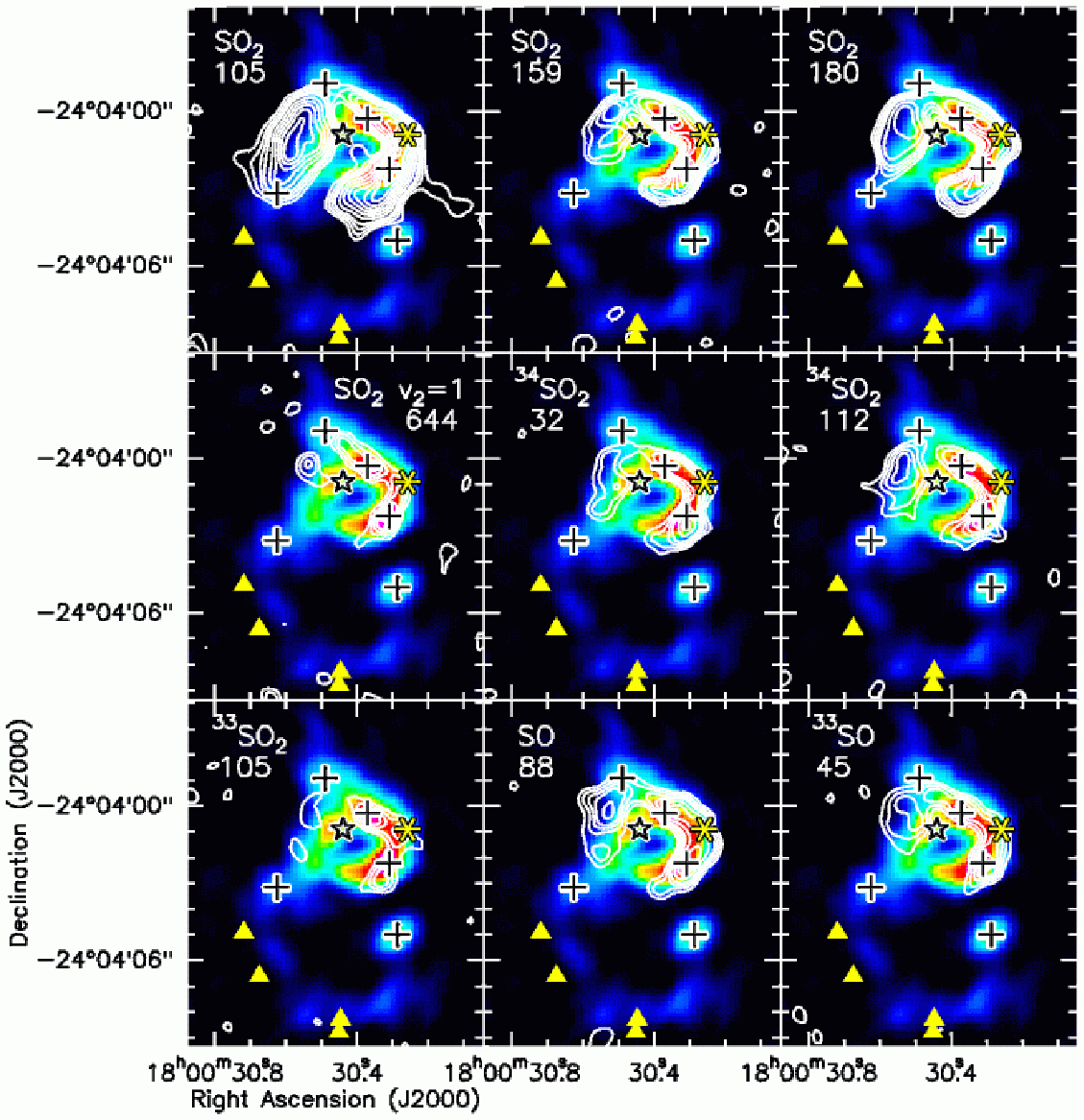} 
\caption{Integrated intensity contour maps of submillimeter SO and \soo\ spectral lines
overlaid on the SMA \smacontwave\ continuum image.  The black crosses
mark the positions of the dust sources SMA1, SMA2, SMA-N, SMA-E, and
SMA-S (see Table~\ref{dustcores}).  The yellow triangles mark the
near-IR H$_2$ knots (A, D1, D2), and the yellow asterisk marks
the origin of the Br$\gamma$ outflow \citep{Puga06}.
Contour levels are 3, 4, 5, 6, 7, 8, 10, 12, 14, 16, 18, 20 $\times$
the base level listed in Table~\ref{smalines}. The number below the
species name is the energy above ground for the lower level in units
of cm$^{-1}$. The image
for \soov\ is the average image of the two lines in the table.
\label{mol9so}}
\end{figure}

Over three dozen spectral lines were detected in our SMA observations
(see Table~\ref{smalines}).  Eighteen of them were detected with a
high signal-to-noise ratio and suffer minimal confusion from line
blending.  To inspect the spatial distribution of emission in these
lines, we created total intensity moment zero images using the total
observed velocity extent of the emission, determined for each line
independently.  Contour maps of these images are shown in
Figures~\ref{mol9} and \ref{mol9so} superposed on the \smacontwave\
continuum image.  In Figure~\ref{mol9so}, the images of two
vibrational lines of \soo\ have been averaged before displaying the
contour map.  In addition, we detect three faint vibrationally-excited
lines of \hcccn\ which we have similarly averaged in order to create
the contour map in Figure~\ref{mol9}.

Figure~\ref{mol9} shows contour maps of a mixture of species,
including high-density tracers, high column-density tracers and
outflow tracers.  As with the dust, none of these gas species show
emission inside the UCHII shell, suggesting a cavity at this location.
Only CO, \cseventeeno, \methanol, \cthirtyfours, and \hthirteencoplus\
exhibit emission significantly (4\arcsec) outside of the UCHII
ring.  Particularly in the southern part of the images, these
molecules appear to be tracing the filamentary structures seen in dust
emission (\S~\ref{continuum}).  In some cases (\methanol\ and
\cthirtyfours), the molecules follow the inner edge of this structure
rather than the structure itself.  Faint emission from
\hthirteencoplus\ and \cseventeeno\ is also found along this
structure.  In contrast, the near-IR H$_2$ features A, D1, and D2
\citep{Puga06} appear along the outer boundary of this structure.  The
strongest emission from \hcccn, \cseventeeno, and \hthirteencoplus\
traces a ridge along the northern and western edges of the UCHII.  The
latter molecule peaks a fraction of an arcsecond further outward and
is the best tracer of the boundary of the mid-infrared emission.  In
addition, these three molecules show an extension toward the
north-northeast which appears to be associated with the dust source
SMA-N.  Other molecules also appear in this vicinity, including
\methanol\ (also seen at this location by \citet{Sollins04}), SiO, and
\cthirtyfours.  This emission is fairly broad ($\Delta v \sim
12$~\kms).  An additional point source of gas emission very close to
the location of the dust source SMA-E is seen in \hthirteencoplus,
\hcccn, \methanol, and \cseventeeno.  Compact peaks of \cseventeeno\
and \hcccn\ are also seen at the position of SMA2.

\begin{figure}
% generated by  sma_lines/cubes/g589_3spec.greg
\includegraphics[height=6.5in,angle=-90]{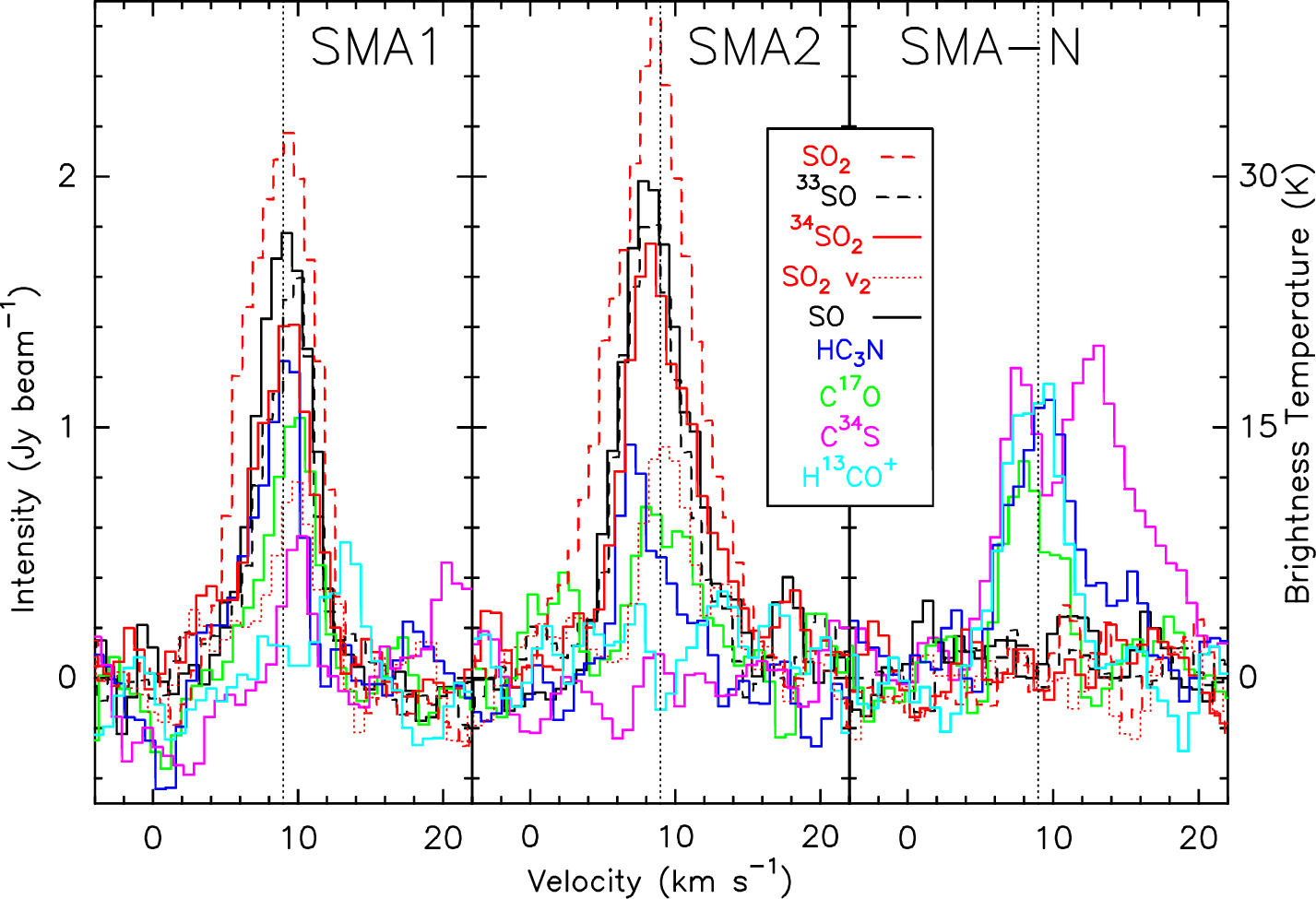} 
\caption{Spectral line profiles extracted from a single data cube pixel 
at the positions of the dust continuum sources SMA1, SMA2, and SMA-N.
The dotted vertical line marks the LSR velocity of +9~\kms. The \soo\ line  
is $16_{7,9}-17_{6,12}$, the \thirtyfoursoo\ line is $17_{4,14}-17_{3,15}$, 
the \hcccn\ line is 38-37, and the \methanol\ line
is $7_{1,7,+0}-6_{1,6,+0}$.
\label{3spectra}}
\end{figure}

Figure~\ref{mol9so} shows integrated intensity contour maps of
emission in nine lines of SO, \soo\ and their isotopologues in our SMA
bandpass.  On the east and west sides of the UCHII ring, the peak
emission from the most abundant species appears to trace ridges that
form a border {\it outside} of the free-free emission.  The position
angles of these ridges are -24\arcdeg\ and -26\arcdeg\ on the east and
west sides, respectively.  In contrast, on the north side, the line
peaks {\it coincide} with the ridge of free-free emission.  Toward the
southeast side, there is no line emission, further suggesting an
incomplete shell in this direction.  The less abundant isotopologues
show peaks at the positions of SMA1 and SMA2, as do the
vibrationally-excited lines of \soo\ and \hcccn\ (Fig.~\ref{mol9}),
indicating high column densities of warm gas.  The
vibrationally-excited lines also show an isolated peak $\sim$1\arcsec\
east-northeast of Feldt's star.  The spectra toward SMA-N are
significantly broader than toward SMA1 or SMA2 and the line wings in
\hcccn\ and \cthirtyfours\ extend to redder velocities (see
Figure~\ref{3spectra}).  This is the same velocity shift detected in
lower angular resolution data, including the OVRO observations of
\chthreecn\ \citep{Akeson96} and the SMA observations of SiO (5-4)
\citep{Sollins04}.

\begin{figure}  
% moment one maps   generated by sma_lines/mom1_6.greg
\includegraphics[height=7.0in,angle=0]{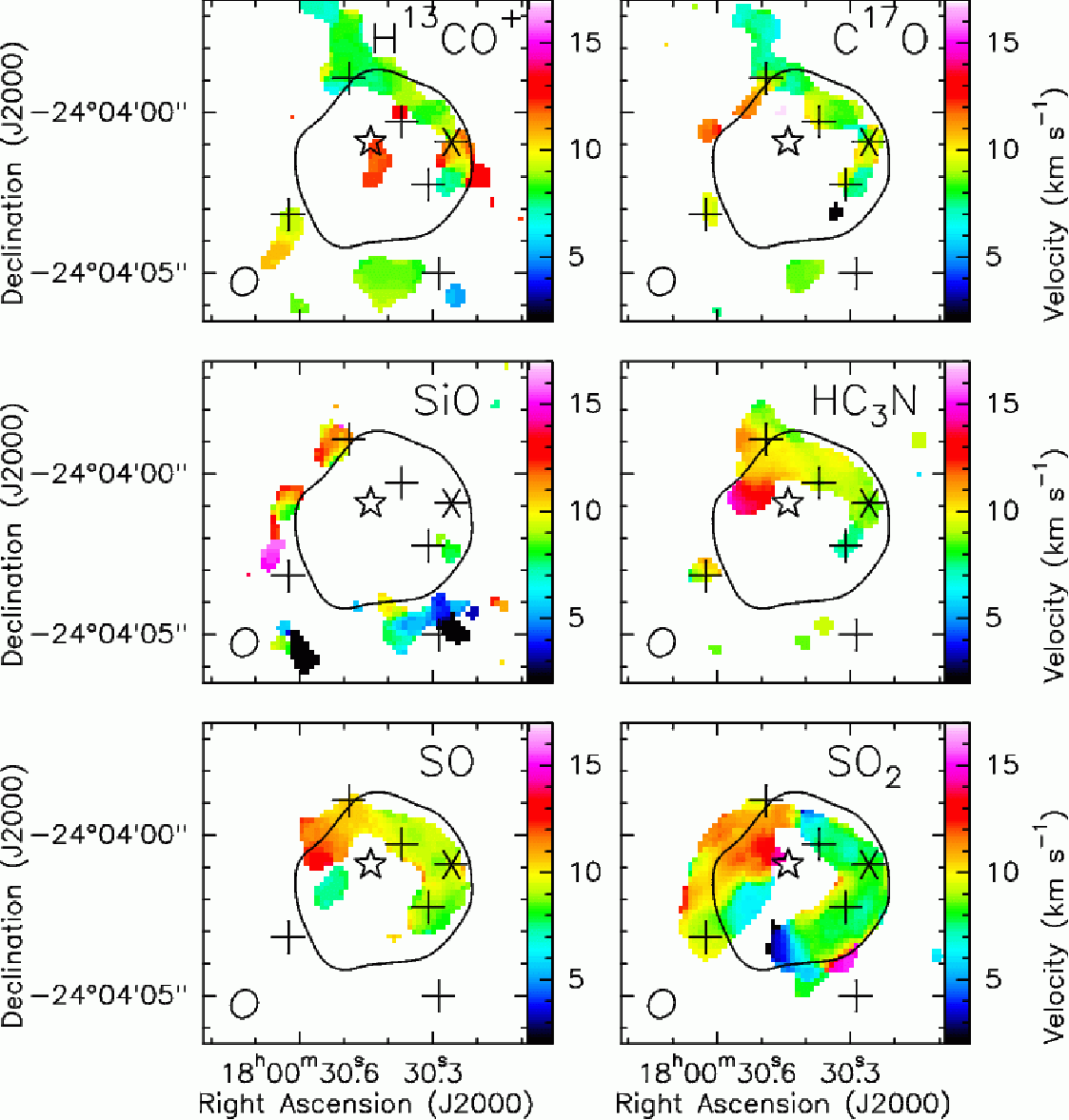} % {f8.ps} 
\caption{Plotted in color scale are the first moment maps of various
  transitions from the \smacontwave\ SMA dataset.  The black contour
  is from the \carmawave\ CARMA continuum image (0.1 \jyb).  The
  crosses mark the dust sources from Table~\ref{dustcores} and the
  asterisk marks the origin of the Br$\gamma$ outflow
  \citep{Puga06}. The star symbol marks the position of Feldt's star
  \citep{Feldt03}.  The SMA beam (\smabeam\ at P.A.=\smapa) is drawn
  at the lower left.
\label{mom1_6}}
\end{figure}

\begin{figure}
% generated by   sma_lines/sollinslines.greg
\includegraphics[width=6.5in,angle=0]{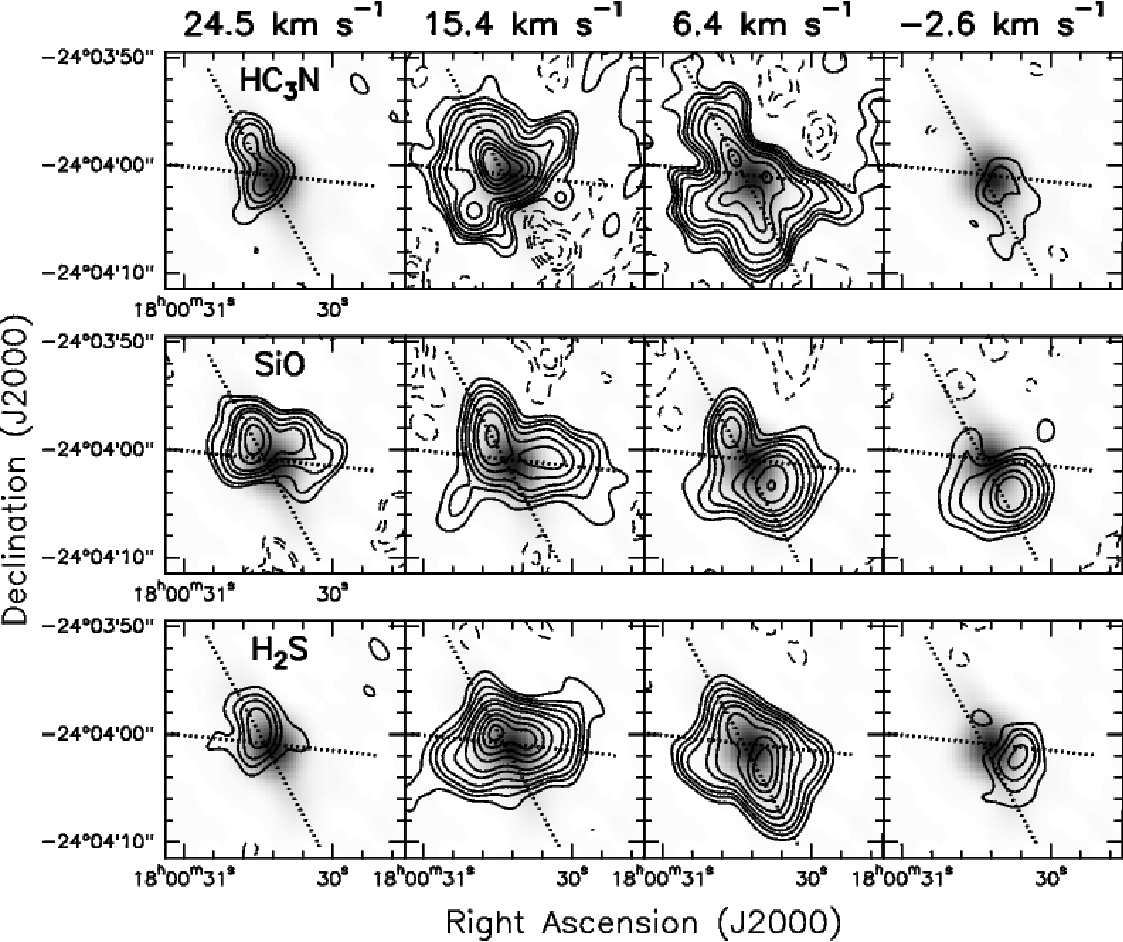} 
\caption{Contour channel maps of three spectral lines from 
the \smammwave\ SMA dataset: \hcccn\ 25-24, SiO 5-4, and 
\htwos\ $2_{2,0}-2_{1,1}$, 
observed with a beamsize of $3\farcs1 \times 2\farcs6$ at P.A.=+5\arcdeg.
Each panel is comprised of eight channels of width 1.1~\kms\ centered 
at the velocity listed; there are no gaps in velocity coverage between 
panels.
Contour levels are -7, -5, -3, 3, 5, 7, 10, 15, 20, 30, 40, 50, 60
$\times 0.6$ \jkms\ for \hcccn, $\times 1.2$ \jkms\ for SiO, and 
$\times 0.55$ \jkms\ for \htwos. 
In each panel, the greyscale image is the 217~GHz continuum, and the
dotted lines mark the position angles reported 
(counter-clockwise from north) in SiO (5-4) \citep{Sollins04}
and CO (1-0) \citep{Watson07}.  The position angle of the 
single-dish CO (3-2) outflow \citep{Klaassen06} 
is similar to that shown for CO (1-0).
\label{sollinslines}}
\end{figure}

The molecular gas shows complicated velocity structure in many of the
species.  Figure~\ref{mom1_6} shows first moment maps of several
species.  Perhaps best seen in \soo, the primary feature is a change
from redshifted emission on the northeast side of the UCHII to the LSR
velocity ($\sim 9$ \kms\/) emission on the northwest and southwest
sides.  The magnitude of the shift is approximately 4~\kms.  The gas
located in the filamentary dust structure near the southern edge of the
images lies at the LSR velocity.  A blueshifted component is seen in
\soo\ at the southern edge of the UCHII.  The \methanol\ and
\cthirtyfours\ show redshifted emission toward SMA-N (see
Fig~\ref{3spectra}).  Redshifted emission is also seen in SiO (8-7)
near SMA-N, while blueshifted emission is seen in the vicinity of
SMA-S.  At lower angular resolution these two spatial components were
interpreted as a bipolar outflow by \citet{Sollins04}.  Although these
areas of SiO emission lie along an axis including Feldt's star, the
emission does not extend back to it.  Having seen these results, we
reanalyzed the lower frequency SMA data from \citet{Sollins04} and
present channel maps of three lines in Figure~\ref{sollinslines}.
Clearly, the SiO (5-4) emission is more complicated than a single
spatial velocity gradient at position angle +28\arcdeg.  In fact, the
emission in all three of these lines demonstrate the complexity of the
molecular emission within $\pm15$~\kms\ of the systemic velocity.  One
is thus driven to studying higher velocity gas in order to seek the
origin of bipolar outflow emission in this region.

\begin{figure}  
% CO moment map   generated by plots/co32hco10sio54.greg   %co32_hco10.greg
\includegraphics[height=3.75in,angle=0]{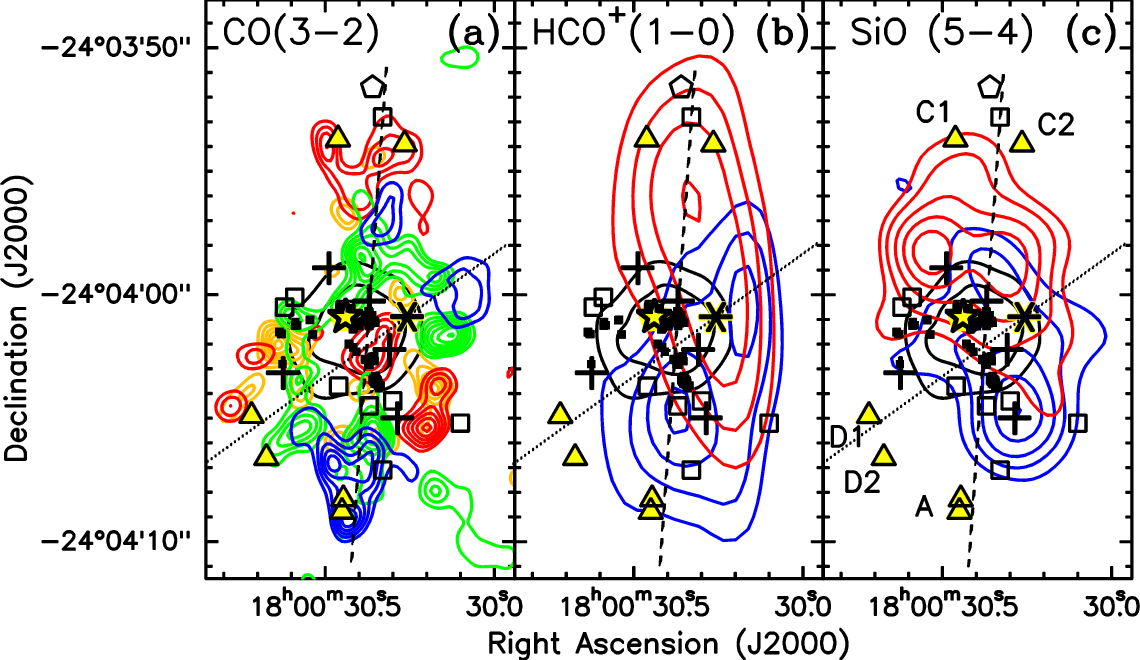} 
\caption{(a) Contour map of CO (3-2) emission observed with a beamsize of 
$0\farcs88\times0\farcs77$ with P.A.=-37\arcdeg.  The velocity
ranges of integration are +80 to +40 \kms\ (red contours), +40 to +20
\kms\ (orange), -4 to -41 \kms\ (green), -41 to -81 \kms\
(blue).  The contour levels are 15, 20, 25, 30, 37.5, 45, 52.5, 60,
67.5 \jkms. (b) Contour map of the high velocity \hcoplus\
(1-0) emission from the dataset of \citet{Watson07} with a beamsize of
$8\farcs5\times3\farcs5$ (P.A. = 1.4\arcdeg); red contours: +30 to
+55~\kms\ with levels 4.5, 6.3, 8.1, 9.9~\jkms; blue contours: -20 to
-45~\kms\ with levels 2, 2.75, 3.5, 4.25~\jkms.  (c) Contour
map of the high velocity SiO (5-4) emission from the 
\citet{Sollins04} data with a beamsize of $3\farcs1 \times 2\farcs6$ at
P.A.=+5\arcdeg, and identical velocity ranges as for \hcoplus.  Levels
are 4, 7, 10, 14, 18 \jkms.  In all panels, the black contours are
from the \carmawave\ continuum image (0.1 and 0.5 \jyb).  Filled
squares are OH masers from \citet{Stark07}, open squares are \water\
masers from \citet{Hofner96}, triangles are the near-IR H$_2$ knots
from \citet{Puga06}, and the pentagon is a Class I \methanol\ maser
position (component 2 from \citet{Kurtz04}).  The dashed line
indicates the outflow axis associated with the dust source SMA1 while
the dotted line represents the orientation of the Br$\gamma$ outflow
and the asterisk marks it origin \citep{Puga06}.
\label{coxmom1}}
\end{figure}

The high abundance of the CO molecule provides the best chance to
study the high velocity molecular gas.  Contour maps showing the
kinematics of CO (3-2) are shown in Figure~\ref{coxmom1}.  In the
midst of the complexity of this emission, we identify a linear bipolar
structure at high velocities at position angle -4\arcdeg.  The axis
defined by this structure passes through the dust core SMA1 and
intercepts highly redshifted CO emission north of the UCHII along with
a Class~I methanol maser (component 2 of \citet{Kurtz04}).  Two
redshifted clumps of CO emission correspond in position to the
redshifted knots C1 and C2 of the near-IR H$_2$ emission
\citep{Puga06}.  A highly redshifted (+78~\kms) water maser feature
\citep{Hofner96} also lies along this axis near the far northern
extent of the CO emission.  Toward the south, the blueshifted knots of
the near-IR H$_2$ emission (component A) lie near the peak of the
blueshifted CO.  Water masers are also found along the southern
portion of the bipolar axis, including a moderately blue-shifted
component at $-3.2$~\kms\/ at the southernmost maser spot.  A higher
velocity blueshifted component was once seen at -61 \kms\ in
single-dish observations by \citet{Genzel77} but has not been detected
in subsequent VLA observations, possibly due to variability.  Most of
the known OH masers, specifically the collections of spots denoted
G5.89~Center and G5.89~South \citep{Stark07}, are clustered in close
proximity to the southern axis of this north-south bipolar structure
and predominantly exhibit the appropriate highly blueshifted
velocities ($\sim$30-40~\kms\ from the LSR).  A notable exception is
the southernmost group of masers in G5.89~South which occur the
southern edge of the UCHII region and have a small redshifted velocity
(from 11-15~\kms).  At the same position and velocity, we see compact
emission in the lowest energy \soo\ line, which appears as a peculiar
red feature in the first moment map (Figure~\ref{mom1_6}).  The other
collection of OH masers, G5.89~East, lies furthest from the bipolar
axis and most of its members have velocities near the LSR, suggesting
that it is unrelated to the north/south structure.

To further explore the north/south bipolar structure, we reimaged and
analyzed the data from the Berkeley-Illinois-Maryland Array (BIMA)
published by \citet{Watson07}.  Of the four molecular transitions
observed, the \hcoplus\ (1-0) line had the widest velocity coverage
(131~\kms).  With a resolution of $8\farcs5\times3\farcs5$, we find
the same north-south velocity gradient in \hcoplus\ emission as is
seen in our CO (3-2) data and the maser data. Contour maps of the red
and blueshifted BIMA \hcoplus\ emission are also shown in
Fig~\ref{coxmom1}.  The sense and direction of the velocity gradient
in CO (3-2) and \hcoplus\ match that seen in a \cthirtyfours\ (3-2)
map obtained with a 17\arcsec\ beam at the IRAM 30m telescope
\citep{Cesaroni91}.  Finally, in Figure~\ref{coxmom1} we have also
plotted the highest velocity emission from the SMA SiO (5-4) data,
extending beyond the range shown in the channel maps in
Figure~\ref{sollinslines}.  Again, the velocity gradient follows a
position angle close to the other lines.

\subsection{\ammonia\ (3,3) maser emission}

\begin{figure}  
% CO 1-0 plus nh3 maser   generated by plots/co10.greg
\includegraphics[height=6.5in,angle=-90]{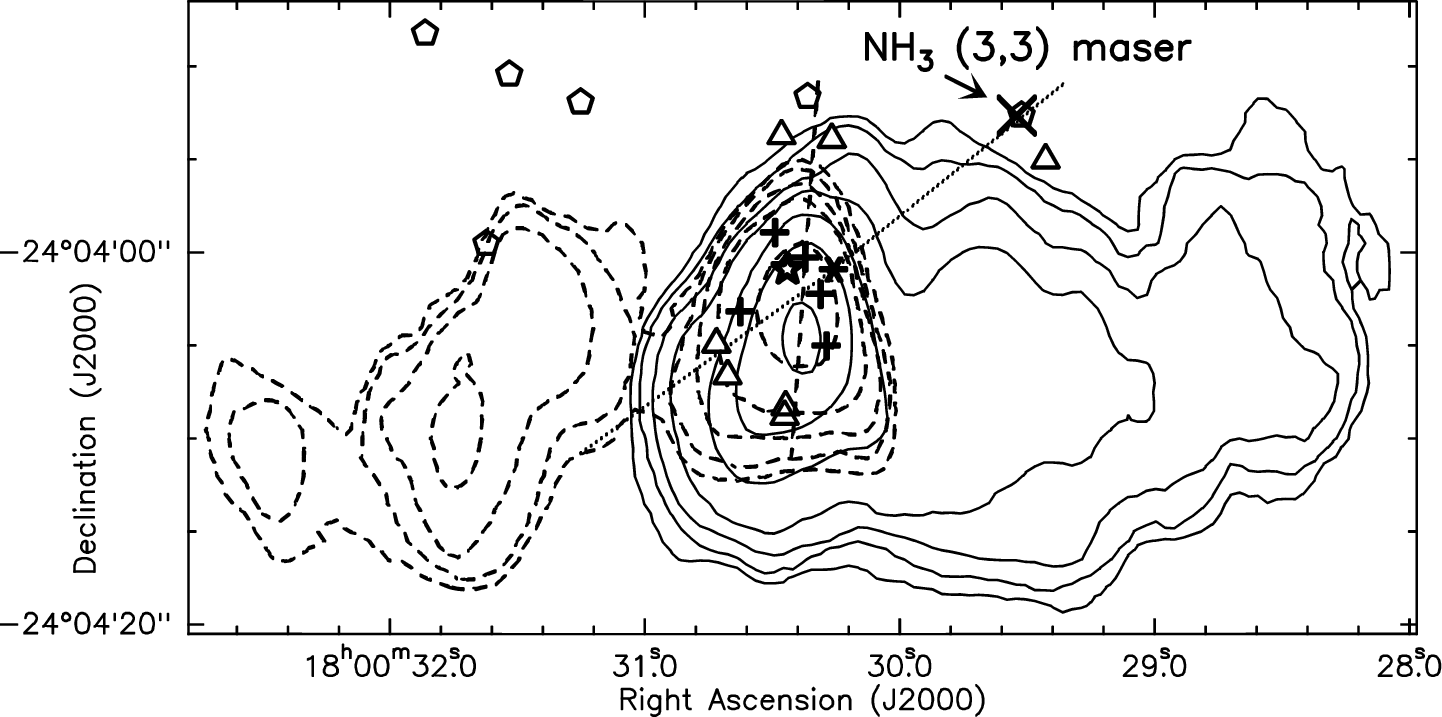}  
\caption{Contour maps of CO (1-0) emission from the dataset of
\citet{Watson07} observed with a beamsize of $7\farcs7\times3\farcs5$
with P.A.=+1\arcdeg.  The velocity ranges of integration are +20 to
+58 \kms\ (dashed contours), and 0 to -37 \kms\ (solid contours).
These ranges begin at slightly higher velocities and extend to higher
velocities than those shown in \citet{Watson07}.  Levels are 5, 10,
20, 40, 80, 160 \jkms.  Triangles are the near-IR H$_2$ knots from
\citet{Puga06}, pentagons are Class I \methanol\ masers from
\citet{Kurtz04}, crosses are the \smacontwave\ dust sources, and the
``X'' is the \ammonia\ (3,3) maser from this work.  The dashed line
indicates the outflow axis associated with the dust source SMA1 while
the dotted line represents the orientation of the Br$\gamma$ outflow
source (asterisk) \citep{Puga06}.
\label{co10}}
\end{figure}

The \ammonia\ (3,3) transition shows a complex arrangement of weak
emission outside the UCHII region and absorption toward the shell,
consistent with the lower resolution data presented by \citet{Gomez91}
and \citet{Wood93}. However, we have discovered an intense point
source of emission located at 18:00:29.539, -24:03:52.65 (J2000),
which is $\approx 12$\arcsec\ northwest of the UCHII region (see
Fig~\ref{co10}). The fitted line peak is at $\sim 11.2$ \kms\/, but
the emission is not spectrally resolved with the 2.45 \kms\/ velocity
resolution of these data (the emission is spread over two
channels). After correction for primary beam attenuation and using
uniform weighting to construct the image, the peak flux density is
$31.3\pm3.0$~\mjb, corresponding to a lower limit for the peak
brightness temperature of $808\pm80$~K at a resolution of
$0\farcs52\times 0\farcs23$.  This temperature is significantly higher
than expected for thermal excitation and indicates weak population
inversion.  Other examples of objects where this \ammonia\ transition
is found to be inverted include the high-mass star-forming regions
DR21(OH) \citep{Mangum94}, NGC~6334 \citep{Kraemer95,Beuther07}, W51
\citep{Zhang95}, and IRAS~20126+4104 \citep{Zhang99}.  In all cases,
these masers are located in or at the ends of high velocity
protostellar outflows.  In \gfive, the peak position and velocity of
the \ammonia\ maser are coincident (to within $0\farcs27$ and
1.0~\kms) with a Class I \methanol\ maser observed with the VLA
(component 1 of \citet{Kurtz04}), which is another outflow tracer.
\gfive\ is only the second object in which these two maser transitions
have been found to coincide spatially, the other example being
DR21(OH) \citep{Mangum94}.  Interestingly, the \ammonia\ maser is also
located within 3\arcsec\ (north-northeast) of the near-IR knot B of
\citet{Puga06}.  As shown in Figure~\ref{co10}, these objects lie
close to the axis (position angle = -54\arcdeg) of the Br$\gamma$
outflow identified by \citet{Puga06} as do the near-IR knots D1 and
D2.  Figure~\ref{co10} summarizes these new results within the context
of the CO (1-0) outflow map from \citet{Watson07}.

\section{Discussion} 

In order to interpret our unprecedented subarcsecond submillimeter
images, we have assembled a comprehensive dataset on this object.  The
multiwavelength nature of our work warrants a fresh examination of the
morphology of the UCHII region, the nature of the ionizing source, the
number and structure of the outflow(s), and the presence of young
stellar objects in this important region.

\subsection{Shape and structure of the ionized nebula}
\label{d1}

UCHII regions are divided into a number of morphological classes
\citep{Wood89}, with one of the rarer classes being ``shell-type''.
\gfive\ is perhaps the most distinctive and well known member of this
class with a well-defined central cavity with a radius approximately
one-third that of the outer UCHII radius.  Our new subarcsecond
submillimeter observations have revealed that this cavity is devoid of
dust, as would be expected when a strong ionizing source is present
within the shell \citep{Churchwell90,Faison98}.  A striking result
from our molecular line images is that, at subarcsecond scales, no
species are seen on the line of sight to Feldt's star, indicating a
low column density of obscuring material close to the star.  This
result is a vivid demonstration of why this star can be seen at
near-IR wavelengths.  In the discovery paper, it was noted that
Feldt's star is significantly offset from the center of the cavity,
and the possibility of proper motion as an explanation was
discussed. At present, there is no knowledge of the velocity of the
star with respect to the UCHII region or the molecular gas.  Another
explanation that may apply to \gfive\ comes from recent modeling of
the dynamical expansion of UCHII regions in self-gravitating molecular
clouds, which demonstrate that nearly spherical shells can be produced
even if the ionizing star is off center \citep{MacLow06}.  In any
case, our data provide no counter evidence to the hypothesis that
Feldt's star is the ionizing source of the UCHII region.

Now we turn our attention to the cm emission just outside of the
shell.  Ever since the original VLA A-configuration 6~cm image
\citep{Zijlstra88}, the elongation of the low-level cm halo of \gfive\
has been evident along a position angle of $\approx -28$\arcdeg.  In
our SMA spectral line images, we now see that the distribution of the
molecular gas traced by \soo\ follows this same position angle as it
effectively encompasses and bounds the UCHII region on two sides (see
the upper left panel of Fig.~\ref{mol9so}).  The morphology of this
line emission suggests that the molecular gas is being driven outward
by the expansion of the UCHII region.  The velocity shifts seen in the
first moment maps (Fig~\ref{mom1_6}) suggest that 2~\kms\ is a lower
limit to the expansion rate (because an additional component of motion
could be present transverse to the line of sight).  The fact that the
molecular gas on the northwestern side of the UCHII peaks on the shell
rather than outside of it, in contrast to the northeast and southwest
sides, suggests that molecular gas is missing from the northwestern
``cap''.  This arrangement would explain the position angle of the
northern elongation of the cm halo.  The possibility that the
ionization front is also ``blowing out'' in the southeastern direction
was previously suggested by \citet{Ball92}.  Our observations support
this suggestion in two respects: we see a break in the \smacontwave\
continuum emission at the point where this southeastern axis
intersects the UCHII shell, and we see extended source of dust
emission (SMA-E) located radially outward from this break.  The fact
that nearly all of the dust emission lies within a few arcseconds of
the outer edge of the UCHII region is a result that agrees with
predictions from previous infrared observations \citep{Harvey94}.

\subsection{Location of Molecules}

As seen in single dish spectra \citep{Thompson99}, we find \gfive\ to
be almost completely lacking in organic molecular line emission while
being rich in sulfur-bearing molecules.  This result is in sharp
contrast to observations of massive young stellar objects (with hot
cores) with the same spectral setup and angular resolution with the
SMA \citep[see e.g.][]{Brogan07a}.  Based on JCMT line profiles of
\gfive, \citet{Thompson99} suggest that \hcccn\ traces the envelope
surrounding the UCHII while the sulfur and silicon-bearing species
with broad lines trace an outflow.  Our images confirm that \hcccn\
emission is located primarily along the northwestern half of the UCHII
shell, and that much of the SiO and \cthirtyfours\ is found at larger
distances from the UCHII.  However, as described in sections~\ref{s1}
and \ref{d1}, we find that SO and \soo\ are also concentrated
primarily around the shell, apparently constraining the expansion of
the UCHII region.  They do not show any obvious evidence of tracing
the beginnings of the larger scale east/west outflow reported in CO
\citep{Klaassen06,Watson07}.  Instead, the optically-thin
isotopologues of SO and \soo\ appear to be tracing compact regions of
emission at the LSR velocity, similar to \cseventeeno\ and \hcccn. At
least two of these compact regions of line emission are coincident
with the dust cores SMA1 and SMA2, while a third may be associated
with the \brgamma\ outflow source described by \citet{Puga06} (see
Figures~\ref{mol9} and \ref{mol9so}).

East of SMA-S, there is an arc-like structure seen in CO, \methanol,
and \cthirtyfours\ just north of a similar filamentary structure seen
in dust emission.  Moreover, emission from the high column density
tracers \cseventeeno\ and \hthirteencoplus\ coincides with the dust
emission.  It is unclear whether this morphology is associated with
the expansion of the UCHII region or is an independent structure.
Remarkably, SMA-S itself is nearly free of line emission in our data.
We find this to be an enigmatic source, similar to NGC6334I~SMA4
\citep{Hunter06}.  It could be a very young protostellar object in
which most common gaseous species are frozen onto grains.  Following
the method described in \S~\ref{sma1sma2}, we estimate the total mass
of SMA-S to be 12-20~\msun\ assuming the temperature is in the range
of 15-20~K.  This amount of mass is sufficient to be an intermediate
mass pre-protostellar core.  Future high-angular resolution
observations in other species that remain in the cold gas phase longer
in the evolutionary sequence, such as \ntwohplus\ and \htwodplus\
\citep{Flower06}, would be useful in studying this object.

\subsection{The Nature of SMA1 and SMA2}
\label{sma1sma2}

There are a growing number of examples of chemical complexity on
arcsecond scales in regions of massive star formation.  One reason for
this complexity is the superposition of multiple young stellar objects
at slightly different LSR velocities, as in, for example Cep-A~East,
\citep{Brogan07b,Comito07}.  The SMA images of \gfive\ have a spatial
resolution of $<2000$ AU, which is finer than the typical separation
of members of protoclusters \citep{Hunter06}.  Thus we are able to
resolve the emission from individual objects and compare their line
strengths in various chemical species.  The set of spectral line
profiles in Figure~\ref{3spectra} shows that the ratio of the strength
of \hcccn\ to SO and \soo\ is larger in SMA1 than in SMA2, as is the
ratio of \cseventeeno\ to SO and \soo.  The high critical density and
excitation temperature of the submillimeter \hcccn\ transitions makes
them an excellent tracer of warm dense gas, as is seen in many
high-mass star forming regions observed at high angular resolution
\citep{Cyganowski07,Mookerjea07,Wright96}.  Also, the strong detection
of the \cseventeeno\ (3-2) line requires a large column density of
moderately warm gas. The relatively greater strength of these two
lines in SMA1 combined with its detection at $11.9\mu$m, and
association with the north/south bipolar outflow suggests that SMA1
likely contains a protostar. SMA2, being richer in sulfur-bearing
molecules, likely has a greater fraction of its molecular excitation
due to shocks, and it is less clear that it harbors a protostar.  We
have fit the line profiles toward SMA1 and SMA2 with single Gaussians
(Table~\ref{sma1sma2lines}) and find that SMA2 is somewhat broader (by
1~\kms) and bluer (by 0.5~\kms), supporting our conclusions.

Unfortunately, it is difficult to estimate the luminosities of SMA1
and SMA2 due to the lack of arcsecond resolution images between
20$\mu$m \citep{Feldt99} and \smacontwave.  The $11.9\mu$m detection
of SMA1 cannot accurately constrain the peak of the dust spectral
energy distribution or the dust temperature.  The best hope for
further progress in the near future would be to measure the gas
temperature.  One can imagine constructing a rotation diagram from our
dataset using the multiple transitions from \soo.  However, this
molecule is clearly very optically thick, as its isotopologues rival
the main line in brightness, and they have somewhat smaller
linewidths.  Thus, an optical depth correction is essential, but this
can only be done using a pair of transitions from the same energy
level.  Although we do have one such pair (\soo\ $19_{1,19}-18_{0,18}$
and \thirtythreesoo\ $19_{1,19}-18_{0,18}$), the comparison is
complicated by the hyperfine structure of \thirtythreesoo.  Accurate
temperature measurements of SMA1 and SMA2 will require a more
extensive combination of spectral lines and isotopologues than are
found in the present SMA data.  In the meantime, a lower limit to the
temperature can be obtained from the brightness temperature of the
brightest line (in both cases it is an \soo\ line).  For SMA1, this
value is 34~K, and for SMA2 it is 65~K.  When compared to the
brightness temperatures of the dust emission in Table~\ref{dustcores},
these values yield upper limits to the \smacontwave\ dust optical
depth of 0.066 and 0.10.  Using these values along with the dust
continuum flux densities, one can derive an upper limit estimate for
the mass.  In the case of SMA1 and SMA2, we use the peak flux
densities in order to assess the unresolved point source component at
these positions.  Following Equation~1 of \citet{Brogan07b}, in which
$\kappa_{875{\mu}m} = 1.84$~cm$^2$~g$^{-1}$ and the gas to dust mass
ratio is 100, we find upper limits of 6~\msun\ and 2~\msun\ for SMA1
and SMA2, respectively. For comparison, if the dust temperature is
100~K, the corresponding masses would be 1.6~\msun\ and 1.3~\msun.

Our observations of \cseventeeno\ (3-2) provide an independent
estimate of the mass of SMA1 and SMA2.  \cseventeeno\ has a well-known
abundance with respect to H$_2$ of $4.7 \times 10^{-8}$
\citep{Frerking82} and has been used in single dish studies of other
ultracompact HII regions \citep{Hofner00}.  At the moderately warm
temperatures of SMA1 and SMA2, the possible confounding effect of
depletion of \cseventeeno\ onto grains should be minimal, as the
desorption temperature is likely in the range of 15-40~K
\citep{Jorgensen06,Doty04}.  We have measured the integrated intensity
of \cseventeeno\ towards these objects, and used the equations of
\citet{Mangum06} to determine the column density in the
optically-thin, Rayleigh-Jean limit.  Using the peak brightness
temperatures of the line, we estimate the optical depths to be small
as long as the excitation temperature is above $\sim 30$~K.  In
Table~\ref{c17o}, we list the opacity, column density and mass
(0.9-1.6~\msun) computed for excitation temperatures of 75~K and
150~K.  The column density and mass have been corrected for opacity.
The values at 75~K are only about 20\% higher than the minimum values
one would obtain if the excitation temperature was set equal to the
upper state energy (32.3~K). A temperature of 100~K provides good
agreement between the dust-derived and \cseventeeno\ circumstellar
mass, and also provides a greybody model consistent with the
submillimeter and mid-IR flux densities at the position of SMA1.
Indeed, based on mid-infrared images at 11.7 and 20~$\mu$m,
\citet{Feldt99} derive a temperature of 120~K for the hot dust
component at the position of SMA1. Using a temperature of 100~K, the
total luminosity of SMA1 is $\sim 3000$~\lsun. For reference, on the
main sequence this luminosity corresponds to an early B star with a
central stellar mass of 7.5 to 8.5~\msun\ depending on the
mass-luminosity relation used \citep{Demircan91,Hilditch87}.

In any event, the circumstellar masses we obtain for SMA1 and SMA2 are
comparable to the upper limits measured for the BN and IRc2 objects in
Orion \citep{Eisner06}, and to circumstellar masses surrounding
intermediate mass protostars \citep{Beltran07,Neri07}.  However, we
emphasize that the nature of SMA2, including whether or not it
contains a central heating source, remains unclear, particularly
because it lacks a clear association with a bipolar outflow (unlike
SMA1).  The locations of SMA1 and SMA2 with respect to the UCHII
region shell and their gas masses are broadly consistent with simulations
of ``secondary collapse'' by \citet{MacLow06}. In the \citet{MacLow06}
model, gravitational instabilities in the material swept up in an
expanding UCHII region shell can produce a second generation of
collapsing cores in the shell with masses of a few \msun. Quantitative
predictions for the dust emission from such cores are not made, but
the prediction for the free-free emission that would be present at
the boundary between the ionized gas and dense core of a few 100 mJy
is consistent with the observed free-free emission in the vicinity of
SMA1 and SMA2. 

The lack of organic ''hot core'' emission from SMA1 (and SMA2 if it
also contains a protostar) has several possible explanations: (1) the
protostar may be of sufficiently late type (low mass) to preclude the
formation of a hot core. However, as already discussed the
circumstellar mass is consistent with those of intermediate mass
protostars, and such sources are capable of producing hot core line
emission \citep[see for example][]{Fuente08}. (2) The protostar could
be sufficiently young that it has not yet reached the hot core phase,
but this seems less likely given the presence of the energetic
north/south outflow from SMA1. (3) The progenitor of the UCHII region
(presumably Feldt's star) may have melted enough of the icy dust
mantles in its vicinity during its formation that the reservoir of
organic material was severely depleted for later generations of
protostars. (4) Although the simulations of \citet{MacLow06} show that
the column densities of the ``secondary collapse'' cores are not
disrupted by the passage of the UCHII shock, the effect on the
chemistry of such cores has not been assessed. Thus it is possible
that the passage of the UCHII shock destroyed the fragile organic
molecules. Higher angular resolution study of the dust emissivity and
molecular line emission with ALMA in the future will help distinguish
between these possibilities.

\subsection{Origin of the outflows}

What do our SMA data reveal about the massive outflow from \gfive?
First of all, the \smacontwave\ line emission shows some evidence of a
general expansion centered on the UCHII region.  The location of the
clump of molecular line emission associated with SMA-N combined with
the ridge of dust, \cseventeeno, and \hthirteencoplus\ emission in the
south suggest that Feldt's star is more likely than SMA1 to be the
origin of this activity.  Regarding the large-scale east/west CO
outflow, there is no obvious bipolar structure tracing back to
Feldt's star or any of the dust cores along the position angles quoted
by \citet{Klaassen06} and \citet{Watson07}.  This result is consistent
with the ``extinct jet'' hypothesis of \citet{Klaassen06}, whereby the
large scale outflow is a remnant flow from a previous generation of
protostellar activity.  However, our CO(3-2) and \hcoplus(1-0) images
do show direct evidence for a collimated bipolar outflow originating
from SMA1 at position angle -4\arcdeg, lending further weight to our
identification of it as a protostar.  This outflow matches the outflow
direction postulated by \citet{Puga06}, as it naturally explains the
arrangement of the near-IR H$_2$ knots A and C.  In addition, this
outflow explains the geometry of most of the maser emission spots in
this region, including \water, OH, and Class I \methanol\ masers.  In
particular, the locations and kinematics of the high-velocity \water\
maser components to the north and south of the UCHII region, along
with the blueshifted OH masers to the south, follow the CO and
\hcoplus\ velocity field centered on SMA1 (see Fig.~\ref{coxmom1}).

In addition to the outflow from SMA1, we also find tentative evidence
for a second outflow in redshifted and blueshifted CO emission and
blueshifted \hcoplus\ emission, centered on the Br$\gamma$ outflow
origin, with a position angle of -54\arcdeg, similar to that found by
\citet{Puga06}.  The near-IR knots to the northwest and southeast (B
and D) are aligned roughly with this axis, as is the new \ammonia\
(3,3) maser.  It seems likely that all of these phenomena are
outflow-related, and have a common driving source.  We find a point
source of emission from vibrationally-excited \soo\ and \hcccn\ at
this position (see Fig.~\ref{mol9} and \ref{mol9so}), suggesting a
central powering source at this location.  While we do not find a compact
dust source at the origin, it does lie at the $\sim 200$~mJy level
within a ridge of dust emission associated with the northwest edge of
the UCHII region.  We estimate an upper limit of $\sim 80$~mJy for a
point source at this position, which corresponds to a mass upper limit
of $\sim 0.4$~\msun\ (assuming a temperature of 65~K), comparable to
gas masses found to be surrounding intermediate-mass protostars
\citep{Beltran07,Neri07}.  Another explanation for knots B and D and
the (3,3) maser is that they are simply enhancements in the
large-scale east/west outflow.  The fact that the \ammonia\ maser (and
its associated \methanol\ maser) sit at the edge of the large
blueshifted CO (1-0) outflow lobe (Figure~\ref{co10}) and emit near
the LSR velocity is consistent with them originating from a velocity
coherent column of shocked gas moving transverse to the line of sight,
which is an ideal geometry for generating strong maser features
\citep[see e.g.][]{Lilj00}.

One question that arises from the work presented here is the relative
ages of the compact north/south outflow from the protostar SMA1, the
large scale east/west outflow, the UCHII region and Feldt's star.  We
can estimate the timescale for the north/south outflow from SMA1.
Using the velocity (w.r.t.\ the G5.89 LSR of 9 \kms\/) and projected
offset of the northern \water\ maser relative to SMA1, (69 \kms\/ and
0.073 pc, respectively), the outflow timescale is $1030~(69/v_{rad}
({\rm km~s}^{-1})(1/{\rm tan}(i))$ yr ($v_{rad}$ is the radial
velocity of the outflow and $i$ is the angle between the line-of-sight
and the outflow direction).  Using the southern CO (3-2) emission, the
timescale is about 1700 years, yielding an average timescale of $\sim
1400$~yr.  For comparison, the timescale for the motion of H$_2$ knot
A is 400~yr \citep{Puga06}.  Although the timescale for the
north/south CO outflow is only a lower limit on the outflow age, it is
a factor of five discrepant from the 7700~yr age reported by
\citet{Watson07} based on BIMA CO~(1-0) data. However, the velocity
ranges used by \citet{Watson07} in computing the age do not include
the high velocity gas that is present in the north/south outflow and
are more appropriate to the lower velocity east/west outflow. The
timescale we derive for the north/south outflow is closer to the value
of 2000~yr reported by \citet{Klaassen06}.  This agreement makes sense,
because the highest-velocity CO~(3-2) emission is unresolved in their
single-dish maps, since those spectral channels are completely
dominated by the compact outflow from SMA1 rather than the larger,
older east/west outflow.

The question remains as to the driving source of the east/west flow.
The width of the east/west outflow combined with the confused velocity
field within 15 \kms\ of the LSR prevents any constraint on the
outflow origin to better than a few arcseconds.  As an additional
source of uncertainty, if the driving source has a small relative
motion of 1~\kms, it could have moved $\sim0\farcs8$ during 7700~yr.
Two possible candidates are Feldt's star and SMA2.  Lacking direct
evidence for an internal heating source in SMA2, we find the most
natural scenario for the east/west outflow is that it was driven by
Feldt's star prior to the creation of the UCHII region.  The
expansion of the UCHII region has disrupted the velocity field around
it making it impossible to trace this outflow back to its origin.  For
the same reason, there could be additional bipolar outflows present
(e.g. from other embedded YSOs) that are difficult to discern in this
complex velocity field.

Finally, a possible clue regarding the relative age of the outflow
from SMA1 and the UCHII region may be found in the fact that the
southern lobe of the outflow only appears distinctly in the CO (3-2)
contour maps in the regions outside the boundary of the UCHII region.
One could argue that the outflow from SMA1 predates the UCHII region,
which has recently expanded and disrupted the inner portions of the
bipolar morphology, leading to the OH masers.  Alternatively, if the
UCHII region is significantly older than the proposed 600~yr age
\citep{Acord98} and SMA1 is located behind it, then the southern lobe
of the outflow from SMA1 may have recently drilled through the UCHII
region, creating the OH masers on the front side, and continuing on
south of the UCHII region.  In this picture, the absence of high
velocity redshifted OH features north of SMA1 would be explained by
the high optical depth of the UCHII region at 1.6~GHz.  We believe the
former picture seems more physically plausible, given the lack of a
disturbance in the ionized gas morphology along the outflow direction
(i.e. the position angle of the low-level centimeter emission differs
from the outflow by 24\arcdeg).  In either case, our observations 
provide further evidence that the OH masers in \gfive\ are
associated with a protostellar outflow, a conclusion originally
reached by \citet{Zijlstra90}.  Other examples of this correlation
have been found, for example the TW object of W3(OH) \citep{Argon03}.
Proper motion measurements of the OH masers in \gfive\ should provide
valuable insight on this phenomenon.

\section{Conclusions}

Our subarcsecond submillimeter images of the ultracompact radio source
G5.89-0.39 (W28~A2) have shed new light on this enigmatic source.  By
using a comprehensive set of lower-frequency images, we have modeled
and removed the free-free emission and find five residual sources of
dust emission.  With no dust emission located inside the shell, our
observations support the previously-proposed picture of a dust-free
cavity located inside a shell-like UCHII region with warm,
high-density gas and dust tracing its periphery.  Two of the compact
dust objects, SMA1 and SMA2, exhibit compact spatial peaks in one or
more of the optically thin tracers $^{33}$SO, $^{34}$SO$_2$, and
\cseventeeno.  In CO (3-2) emission, we have identified a
well-collimated, high-velocity outflow from SMA1 at position angle
-4\arcdeg\ and conclude that it is an embedded intermediate-mass
protostar, surrounded by $\approx 1$~\msun\ of circumstellar material
at a temperature of approximately 100~K, and corresponding to the
brightest source in the \mirlambda\ image.  The outflow from SMA1
explains much of the near-IR H$_2$ and centimeter wavelength maser
emission in the region.  We also find tentative evidence for a second
CO outflow associated with the Br$\gamma$ outflow identified by
\citet{Puga06}. The position angle of this outflow points toward the
location of a new \ammonia\ (3,3) maser that we have discovered
12\arcsec\ northwest of the UCHII region.  The origin of this outflow
is marked by compact emission in vibrationally-excited transitions of
\soo\ and \hcccn.  Regarding the dust source SMA2, it is unclear
whether it harbors a central heating source or is the result of a
strong shock.  Nonetheless, the masses and locations of SMA1 and SMA2
are broadly consistent with the \citet{MacLow06} model of secondary
collapse in material swept up in the expanding shells of UCHII
regions.  Located outside the UCHII shell, we detect a cold dust
object (SMA-S) that is remarkably free of line emission.  Assuming a
low temperature of 15-20K, we estimate a mass of 12-20~\msun, which
suggests it is likely to be an intermediate mass pre-protostellar
core.  Beyond SMA-S, we detect the beginning of a filamentary
structure of dust and gas emission that extends for several arcseconds
eastward before turning northeast.  Whether this structure is related
in any way to the UCHII region is an open question.

\acknowledgments

We thank C. Watson for providing access to BIMA data on this region.
CJC is supported by a National Science Foundation Graduate Research
Fellowship.  RI was supported in part during this research by a NASA
Spitzer fellowship to UVa.  This research has made use of NASA's
Astrophysics Data System Bibliographic Services and the SIMBAD
database operated at CDS, Strasbourg, France, and is based in part on
observations collected at the European Southern Observatory, Paranal,
Chile.  This research used the facilities of the Canadian Astronomy
Data Centre operated by the National Research Council of Canada with
the support of the Canadian Space Agency.  TRH thanks K. T. Constantikes
for introducing him to the power of Matlab.

%{\it Facilities:} \facility{SMA}, \facility{VLA}, \facility{CARMA}, \facility{JCMT}

\clearpage

\clearpage

\clearpage

\clearpage

\clearpage
 
\clearpage

\clearpage

\clearpage

\clearpage

\clearpage

\clearpage

\clearpage

\begin{table}
\caption{Summary of VLA archival data \label{vlaobs}  }
\scriptsize
\begin{tabular}{ccccccccl}   
\tableline
\tableline
% VLA config. is available in FITS header in HISTORY for FILLM 
        &            &       &            &     &         & Continuum   &                     & \\
Project & Date       & Freq. & Wave. &     &         & sensitivity & Beam                & \\
Code    & YYYY-MM-DD & (GHz) & (cm) & Observation Type & Config. & (\mjb)      & \arcsec\ $\times$ \arcsec\ (P.A.)& Calibrators\\
\tableline
AW158\tablenotemark{a} & 1986-04-27 & 5 & 6.0 & continuum  & A    & 1.3   & 
      $0\farcs85 \times 0\farcs43$ (-17\arcdeg)  & 3C286, J1748-253 \\ % G5.89_CBAND.HGEOM 
AA227 & 1998-06-24 & 8.4 & 3.6 & continuum              & BnA  & 0.17   & 
      \xbandbeam\ (+73\arcdeg) & 3C286, J1730-130 \\ % G5.89_X_R5.HGEOM
AW550 & 2001-05-12 & 15 & 2.0 & recomb. line           & B    & 0.30  &
      \ubandbeam\ (-1\arcdeg) & J2251+158 \\ %G5.89_U76_C.HGEOM
AZ075 & 1995-11-17 & 22 & 1.3 & \ammonia\ (2,2), (3,3) & B    & 1.8\tablenotemark{b}   &
      $0\farcs73 \times 0\farcs32$ (+22\arcdeg) & 3C286, J1730-130 \\ % G5_22_33_C.HGEOM mode 4
%
%AT243 & 2000-03-18 & 22  & \ammonia\ (4,4)        & C    & 0.96  &
%$1\farcs88 \times 0\farcs92$ ($+2^\circ$)  & J1730-130 \\ %G5.89_44_C.HGEOM  mode 1A
%
%AA185 & 1995-08-   & 22  & \water\                & A    &       &  &  \\
%
%AC792\tablenotemark{b} & 2005-10-17 & 43  & SiO (1-0)               & DnC   & 1.6   &     
%$1\farcs34 \times 0\farcs89$ ($+90^\circ$)  & 3C286, J1820-254 \\  %G5.89_QMOSAP.HGEOM
\tableline
\end{tabular} 
\tablenotetext{a}{~Data originally published by \citet{Wood89}.}
\tablenotetext{b}{~Line free channels from both IFs combined.}
\end{table}

\begin{deluxetable}{llllll}
\tablecaption{ESO Archival Infrared Images\label{esodata}}
\tablehead{
\colhead{telescope} & \colhead{instrument} & \colhead{filter/band} &
\colhead{observation date} & \colhead{program} & \colhead{PI}}
\startdata
VLT-Yepun & NACO & H$_2$ 2.122$\mu$m      & 2004/06/15 & 073.C-0178(A) & Feldt \\
VLT-Yepun & NACO & Br$\gamma$ 2.166$\mu$m & 2004/06/15 & 073.C-0178(A) & Feldt \\
VLT-Yepun & NACO & L' 3.8$\mu$m           & 2004/06/15 & 073.C-0178(A) & Feldt \\
VLT-Yepun & NACO & K$_s$ 2.1$\mu$m        & 2004/06/15 & 073.C-0178(A) & Feldt \\
NTT  & SOFI  & Br$\gamma$             & 1999/08/01 & 63.H-0292(A)  & Kaper \\
NTT  & SOFI  & Br$\gamma$-continuum   & 1999/08/01 & 63.H-0292(A)  & Kaper \\
NTT  & SOFI  & 1.215$\mu$m            & 1999/08/01 & 63.H-0292(A)  & Kaper \\
VLT-Antu & ISAAC & 1.71$\mu$m             & 2002/07/09 & 69.C-0448(A) & Bik \\
3.6m & TIMMI2 & 11.9$\mu$m            & 2003/07/20 & 71.C-0342(A) & Brooks \\
\enddata
\end{deluxetable}

\begin{table}
\caption{Comparitive continuum intensities at various positions in the UCHII region \label{cmflux}} 
\begin{tabular}{lcccccc}
\tableline
\tableline
                                   & \multicolumn{6}{c}{Flux density (\jyb)\tablenotemark{a}} \\
Position                           & 6~cm  & 3.6~cm & 2~cm & 1.3~cm & 3~mm & \smacontwave \\
\tableline
SMA1                               & 0.205 & 0.449 & 0.706 & 0.863 & 0.771 & 1.11 \\ % 267,260
SMA2                               & 0.151 & 0.375 & 0.652 & 0.812 & 0.740 & 0.987 \\ % 279,232
center of shell\tablenotemark{b}   & 0.202 & 0.386 & 0.535 & 0.614 & 0.528 & 0.349 \\ % 261,244
$0\farcs3$ east of Feldt's star\tablenotemark{c} & 0.199 & 0.429 & 0.673 & 0.829 & 0.765 & 0.632 \\ % 249,250
total (within a 12\arcsec\ box)        & 2.36  & 4.93  & 7.37  &  8.81 & 7.51  & 6.44 \\
calibration uncertainty            & 5\%   & 5\%   & 5\%   & 10\%  & 10\%  & \smaunc\%  \\ 
\tableline
\end{tabular}
\tablenotetext{a}{All images have been restored to the beamsize of the 3~mm image: \commonbeam\ at P.A. = \commonpa.}
\tablenotetext{b}{J2000 position: 18:00:30.402, -24:04:01.39} % 261,244
\tablenotetext{c}{J2000 position: 18:00:30.463, -24:04:00.97} % 249,250
\end{table}

\begin{table}
\caption{\smacontwave\ positions and flux densities of dust sources \label{dustcores}  }
\begin{tabular}{cccccc}   
\tableline
\tableline
       &   &                         & Peak flux  & Peak brightness & Integrated \\ 
       & \multicolumn{2}{c}{Position} & density\tablenotemark{a} & temperature & flux density\tablenotemark{a,b} \\
Source &  $\alpha$ ($^{\rm h}~~^{\rm m}~~^{\rm s}$)   & $\delta$ ($^{\circ}~~{'}~~{''}$) & \jb & K & Jy \\
\tableline
% 1 beam = 163.38 pixels 
SMA1  & 18 00 30.37 & -24 04 00.27 & $0.50$ & 4.2 & $0.50\pm0.08$\tablenotemark{c} \\ % blc -1,16 trc 268,260 = 801 pix  135,137
SMA2  & 18 00 30.30 & -24 04 02.37 & $0.40$ & 3.3 & $0.40\pm0.06$\tablenotemark{c} \\ % blc -1,14 trc 280,230 = 606 pix  142,123
SMA-N & 18 00 30.48 & -24 03 58.87 & $0.31$ & 2.6 & $0.74\pm0.11$ \\ % blc -1,17 trc 242,282 = 804 pix  125,147
SMA-S & 18 00 30.29 & -24 04 05.03 & $0.38$ & 3.2 & $0.52\pm0.09$ \\ % blc -1,16 trc 283,192 = 654 pix  143,105
SMA-E & 18 00 30.62 & -24 04 03.14 & $0.19$ & 1.6 & $0.45\pm0.08$ \\ % blc -1,16 trc 216,216            113,118
\tableline
\end{tabular} 
\tablenotetext{a}{values taken from the free-free subtracted image with resolution of \commonbeam}
\tablenotetext{b}{uncertainties include the \smaunc\% calibration uncertainty}
\tablenotetext{c}{source is unresolved}
\end{table}

\begin{table}
\caption{Spectral lines detected in the \smacontwave\ SMA data \label{smalines}}
\begin{tabular}{lccccc}
\tableline 
Species    & Transition & Frequency  & Sideband & E$_{\rm lower}$ & Base contour\\
           &            & (GHz)      &          & (cm$^{-1}$)     & \jkms \\
\tableline 
\tableline  
\methanol        & $7_{1,7,+0}-6_{1,6,+0}$ & 335.58200                  & lower & 43.7  & 4.02 \\ % chunk 23 -
\soo             & $23_{3,21}-23_{2,22}$   & 336.08923                  & lower & 180.6 & 3.76 \\ % 18 +
\hcccn           & 37-36                   & 336.52008                  & lower & 202.1 & 2.40 \\ % 12 -
SO               & $11_{10}-10_{10}$       & 336.55381                  & lower & 88.1  & 2.87 \\ % 11/12 -/-
\soo             & $16_{7,9}-17_{6,12}$    & 336.66958                  & lower & 159.1 & 2.76 \\ % 10 +
\methanol        & $12_{1,11,-0}-12_{0,12,+0}$ & 336.86511              & lower & 125.7 & 3.55 \\ % 8  -
\cseventeeno     & 3-2                     & 337.06113\tablenotemark{b} & lower & 11.2  & 1.65 \\ % 5  +
\thirtythreeso   & $8_{7,9}-7_{6,8}$       & 337.19862                  & lower & 44.7  & 3.73 \\ % 4  -
\hcccnv          & $37_1 - 36_{-1}$        & 337.34469                  & lower & 425.6 & 0.90 \\ % 2  + LSB=-freq,+vel
\cthirtyfours    & 7-6                     & 337.39646                  & lower & 33.8  & 4.46 \\ % 1  +
\tableline
\thirtyfoursoo   & $7_{4,4}-7_{3,5}$       & 345.51966                  & upper & 32.7  & 3.30 \\ % 1  +
\thirtyfoursoo   & $6_{4,2}-6_{3,3}$       & 345.55309                  & upper & 28.2  & ..   \\ % 1  +

\thirtythreesoo  & $19_{1,19}-18_{0,18}$   & 345.58500\tablenotemark{a} & upper & 105.1 & 2.81 \\ % 2 + USB=+freq,-vel
\hcccn           & 38-37                   & 345.60901                  & upper & 213.3 & ..   \\ % 2 +
\thirtyfoursoo   & $5_{4,2}-5_{3,3}$       & 345.65129                  & upper & 24.4  & ..\tablenotemark{c} \\ % 2/3 +/-
\thirtyfoursoo   & $4_{4,0}-4_{3,1}$       & 345.67879                  & upper & 21.1  & ..\tablenotemark{c} \\ % 3 -
\twelveco        & 3-2                     & 345.79599                  & upper & 11.5  & 41.4 \\ % 4 -
\methanol        & $16_{1,15}-15_{2,14}$   & 345.90397                  & upper & 219.7 & ..\tablenotemark{c} \\ % 6 +
\methanol        & $18_{-3,16}-17_{-4,14}$ & 345.91919                  & upper & 307.8 & ..\tablenotemark{c} \\ % 6 +
\thirtyfoursoo   & $17_{4,14}-17_{3,15}$   & 345.92935                  & upper & 112.5 & 3.26  \\ % 6 +
\methanol        & $5_{4,2}-6_{3,3}$       & 346.20278                  & upper &  68.5 & ..\tablenotemark{c} \\ % 9 +
NS               & $8_{1,-1,9}-7_{1,1,8}$  & 346.22014                  & upper &  37.8 & ..\tablenotemark{c} \\ % 9/10 +/+
\hcccnv          & $38_{-1} - 37_1$        & 346.45573                  & upper & 436.8 & ..    \\ % 12 -
\soo             & $16_{4,12}-16_{3,13}$   & 346.52388                  & upper & 102.7 & ..\tablenotemark{c} \\ % 13 +
SO               & $8_9-7_8$               & 346.52848                  & upper & 43.2  & ..\tablenotemark{c} \\ % 13 +
\soov            & $18_{4,14}-18_{3,15}$   & 346.59179                  & upper & 643.6 & 1.62  \\ % 14 +
\soo             & $19_{1,19}-18_{0,18}$   & 346.65217                  & upper & 105.3 & 5.16  \\ % 15 -
\hcccnv          & $38_1 - 37_{-1}$        & 346.94912                  & upper & 437.1 &  -    \\ % 18 +
\hthirteencoplus & 4-3                     & 346.99834                  & upper & 17.4  & 3.10  \\ % 19 -
SiO              & 8-7                     & 347.33056                  & upper & 40.6  & 5.77  \\ % 23 -
\tableline  
\end{tabular}
\tablenotetext{a}{comprised of 4 hyperfine components}
\tablenotetext{b}{comprised of 14 components}
\tablenotetext{c}{blended with adjacent lines}
\end{table}

\begin{table}
\caption{Results of Gaussian fits of lines toward SMA1 and SMA2 \label{sma1sma2lines}}
\begin{tabular}{lccccccc}
\tableline
% in the molecular line cubes:
% SMA1 = 136,137
% SMA2 = 141,124
% SMA-E = 113, 118
% BrGamma origin = 146, 133
\tableline
        &                 & \multicolumn{3}{c}{SMA1}  & \multicolumn{3}{c}{SMA2} \\
Species & E$_{\rm lower}$ & Amplitude\tablenotemark{a}    & Velocity       & FWHM           & Amplitude\tablenotemark{a}       & Velocity       & FWHM \\
        & (cm$^{-1}$)     & (\jb) & (km s$^{-1}$) & (km s$^{-1}$) & (\jb) & (km s$^{-1}$) & (km s$^{-1}$) \\ 
\tableline
\thirtyfoursoo  &  32.7 & 1.38$\pm$ 0.06 & 9.06$\pm$ 0.11 & 5.27$\pm$ 0.28 &  1.69$\pm$ 0.08 & 8.64$\pm$ 0.13 & 6.08$\pm$ 0.34 \\
\thirtythreesoo & 105.1 & 1.03$\pm$ 0.06 & 9.32$\pm$ 0.14 & 4.92$\pm$ 0.35 &  1.06$\pm$ 0.06 & 8.82$\pm$ 0.11 & 4.18$\pm$ 0.27 \\
\hcccn          & 213   & 1.17$\pm$ 0.10 & 8.95$\pm$ 0.17 & 3.96$\pm$ 0.42 &  0.83$\pm$ 0.06 & 8.08$\pm$ 0.17 & 4.77$\pm$ 0.45 \\
\thirtyfoursoo  & 112.5 & 1.32$\pm$ 0.06 & 9.30$\pm$ 0.11 & 4.83$\pm$ 0.26 &  1.51$\pm$ 0.06 & 9.12$\pm$ 0.12 & 5.80$\pm$ 0.30 \\
\soo            & 105.3 & 2.06$\pm$ 0.06 & 7.76$\pm$ 0.09 & 5.96$\pm$ 0.22 &  4.33$\pm$ 0.06 & 8.08$\pm$ 0.06 & 7.94$\pm$ 0.14 \\
\soo            & 180.6 & 2.28$\pm$ 0.06 & 8.44$\pm$ 0.07 & 5.69$\pm$ 0.17 &  2.36$\pm$ 0.05 & 8.13$\pm$ 0.07 & 7.22$\pm$ 0.17 \\
\hcccn          & 202.1 & 1.06$\pm$ 0.09 & 9.03$\pm$ 0.17 & 4.00$\pm$ 0.42 &  0.90$\pm$ 0.05 & 8.51$\pm$ 0.10 & 3.47$\pm$ 0.25 \\
SO              & 88.1  & 1.81$\pm$ 0.05 & 9.43$\pm$ 0.06 & 4.42$\pm$ 0.15 &  1.91$\pm$ 0.06 & 8.78$\pm$ 0.08 & 5.37$\pm$ 0.20 \\
\soo            & 159.1 & 1.67$\pm$ 0.05 & 8.59$\pm$ 0.08 & 5.38$\pm$ 0.20 &  2.08$\pm$ 0.07 & 8.17$\pm$ 0.10 & 5.90$\pm$ 0.25 \\
\cseventeeno    & 11.2  & 1.05$\pm$ 0.07 & 9.18$\pm$ 0.13 & 3.99$\pm$ 0.32 &  0.59$\pm$ 0.08 & 8.82$\pm$ 0.29 & 4.54$\pm$ 0.71 \\
\thirtythreeso  & 44.7  & 1.48$\pm$ 0.07 & 9.76$\pm$ 0.09 & 4.10$\pm$ 0.21 &  1.74$\pm$ 0.06 & 8.66$\pm$ 0.08 & 4.80$\pm$ 0.19 \\
\tableline
\multicolumn{2}{l}{weighted average} &   & 8.94$\pm$ 0.03 & 4.94$\pm$ 0.07 &                 & 8.42$\pm$ 0.03 & 6.07$\pm$ 0.07 \\
\tableline
\tablenotetext{a}{The conversion from flux density to brightness temperature is approximately 15.0 K/Jy in both sidebands.}
\end{tabular}
\end{table}

\begin{table}
\caption{Gas masses from \cseventeeno\ (3-2) \label{c17o}}
\begin{tabular}{lcccccccc} 
\tableline
\tableline
       &  Integrated                 &  \multicolumn{3}{c}{$T_{\rm exc} = 75$ K} & \multicolumn{3}{c}{$T_{\rm exc} = 150$ K} \\ 
Source &  Intensity\tablenotemark{a} & $\tau_{\rm C^{17}O(3-2)}$ & N(\cseventeeno)
       & H$_2$ mass\tablenotemark{b} & $\tau_{\rm C^{17}O(3-2)}$ & N(\cseventeeno)    
       & H$_2$ mass\tablenotemark{b} \\
Name   &        (Jy \kms)        & (neper)             & log(\persqcm) & (\msun) 
                                 & (neper)             & log(\persqcm) & (\msun)   \\ 
\tableline
SMA1   & $4.1\pm0.8$             & 0.24 &        12.79 & $1.1\pm0.3$ & 0.11 & 12.97 & $1.6\pm0.4$\\
SMA2   & $3.6\pm0.7$             & 0.13 &        12.70 & $0.9\pm0.3$ & 0.06 & 12.89 & $1.4\pm0.3$\\
%Br$\gamma$ outflow origin & $5.0\pm1.0$ & 0.15 & 12.85 & $1.3\pm0.4$ & 0.07 & 13.04 & $1.9\pm0.5$\\
\tableline
\tablenotetext{a}{Measured at the positions in Table~\ref{dustcores}; includes absolute calibration uncertainty}
\tablenotetext{b}{Assuming the abundance ratio of \cseventeeno\ to H$_2$ to be $4.8\times 10^{-8}$}
\end{tabular}
\end{table}

\end{document}